\documentclass[
reprint,
superscriptaddress,
amsmath,
amssymb,
aps,
prl,
]{revtex4-1}

\usepackage{graphicx}   
\usepackage{comment}    
\usepackage{dcolumn}    
\usepackage{bm}         
\usepackage[
colorlinks=true,
citecolor=MidnightBlue,
linkcolor=BrickRed,
urlcolor=blue,
breaklinks=true,
]{hyperref}   
\usepackage[usenames,dvipsnames]{xcolor}
 
\usepackage{lipsum, babel}
\usepackage{braket}
\usepackage{upgreek}
\usepackage[T1]{fontenc}
\usepackage{mathrsfs}

\DeclareMathAlphabet{\mathsfit}{\encodingdefault}{\sfdefault}{m}{sl}
\SetMathAlphabet{\mathsfit}{bold}{\encodingdefault}{\sfdefault}{bx}{sl}
\newcommand{\tens}[1]{\bm{\mathsfit{#1}}}

\graphicspath{{Figs/}}

\makeatletter
\def\@fnsymbol#1{
	\ensuremath{\ifcase#1\or
	*\or				
	\ddagger\or			
	\dagger\or			
	\mathsection\or		
	\mathparagraph\or	
	\|\or				
	**\or				
	\ddagger\ddagger\or	
	\dagger\dagger		
	\else\@ctrerr\fi}}
\makeatother

\begin{document}

%
%

\preprint{APS/123-QED}

\title{Enhanced dispersion in an oscillating array of harmonic traps}

\author{Joseph M. Barakat}
\email{josephbarakat@ucsb.edu}
\affiliation{
    Department of Chemical Engineering, University of California, Santa Barbara, Santa Barbara, CA 93106
}

\author{Sho C. Takatori}
\email{stakatori@ucsb.edu}
\affiliation{
    Department of Chemical Engineering, University of California, Santa Barbara, Santa Barbara, CA 93106
}

\date{\today}

%
%

\begin{abstract}
	Experiment, theory, and simulation are employed to understand the dispersion of colloidal particles in a periodic array of oscillating harmonic traps generated by optical tweezers.
In the presence of trap oscillation, a non-monotonic and anisotropic dispersion is observed. 
Surprisingly, the stiffest traps produce the largest dispersion at a critical frequency, and the particles diffuse significantly faster in the direction of oscillation than those undergoing passive Stokes-Einstein-Sutherland diffusion.
Theoretical predictions for the effective diffusivity of the particles as a function of trap stiffness and oscillation frequency are developed using generalized Taylor dispersion theory and Brownian dynamics simulations.
Both theory and simulation demonstrate excellent agreement with the experiments, and reveal a new ``slingshot'' mechanism that predicts a significant enhancement of colloidal diffusion in dynamic external fields.
\end{abstract}

%
%

\maketitle

%
%

%
%

The dispersion of colloidal particles in dynamic external fields underlies many transport processes.
Many studies have analyzed the effective diffusivity of particles under a static, external potential \cite{Fulde1975,Festa1978,Das1979,Weaver1979}, including porous media \cite{Brenner1993,Mangeat2020}, block copolymers \cite{Fredrickson1991}, corrugated substrates \cite{Ma2015}, and colloidal crystals \cite{Loudiyi1992,Bechinger2000}.
Experimentally, focused lasers have been used to create two-dimensional (2D) arrays of potential wells to study the freezing and melting of colloidal crystals \cite{Loudiyi1992,Bechinger2000}.
Although passive transport of colloids in a static external field is well studied, many transport processes involve nonequilibrium driving forces that generate a non-trivial coupling between convective and diffusive motion.

%
%

\begin{figure}[!b]
	\vspace{-10pt}
	\centering
	\includegraphics[width=1.0\linewidth]{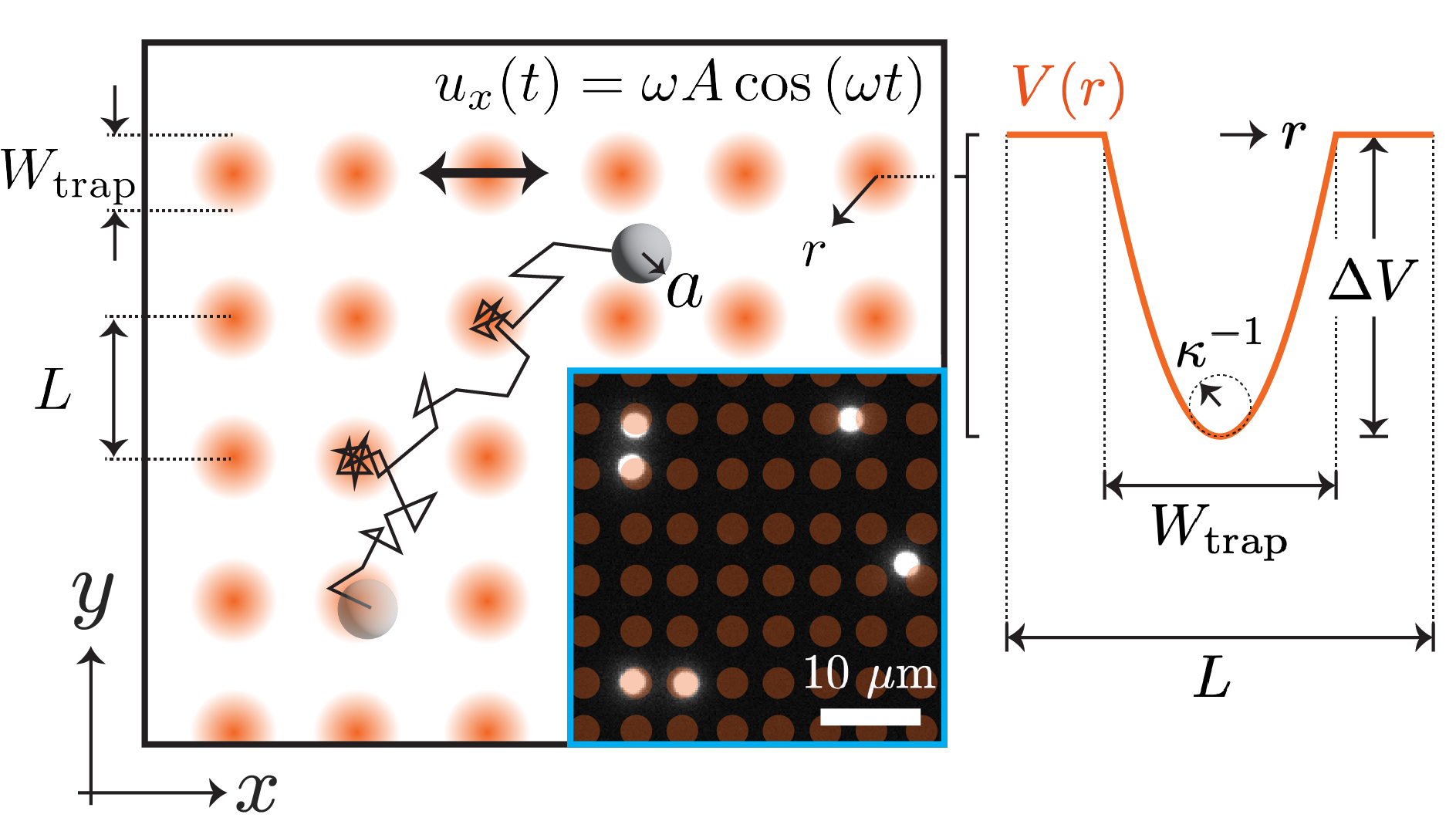}
	\caption{%
		Schematic of a Brownian particle diffusing in a two-dimensional (2D), oscillating array of harmonic traps with potential-energy field $V(\bm{r})$ and velocity $\bm{u}(t)$ given by Eqs.~\eqref{eq:harmonic_potential} and \eqref{eq:trap_velocity}, respectively.
		The harmonic well has curvature $\kappa$ and depth $\Delta V = \frac{1}{8}\kappa W_{\text{trap}}^{2}$.
		Inset: Experiment snapshot of radius $a = 1.25$ $\mu$m silica particles diffusing in an array of traps created by optical tweezers.
	}
	\label{fig:Fig1}
\end{figure}

In this Letter, we combine experiment, theory, and simulation to study the dispersion of colloidal particles in a time-varying array of mobile potential wells.
Experimentally, we use an optical tweezer to generate a 16 $\times$ 16 lattice of harmonic traps spaced a distance $L=6$ $\mu$m apart along a 2D plane (see Fig.~\ref{fig:Fig1} for a schematic of our experimental system).
The interaction of a colloidal particle with each trap is well-modeled by the piecewise potential,
\begin{equation}
	V(\bm{r}) 
	=
	\begin{cases}
		\tfrac{1}{2} \kappa r^{2}
		& \text{for } r \le \tfrac{1}{2} W_{\text{trap}},
		\\
		\Delta V
		& \text{for } r > \tfrac{1}{2} W_{\text{trap}},
	\end{cases}
	\label{eq:harmonic_potential}
\end{equation}
where $\bm{r}$ is the position relative to the trap's center, $\kappa$ is the trap stiffness, $W_{\text{trap}}$ is the trap width ($\approx 3.2$ $\mu$m), and $\Delta V = \tfrac{1}{8} \kappa W_{\text{trap}}^{2}$ is the potential well depth (see the Supplemental Material \cite{Supplemental} for details on quantifying these parameters).
Most optical tweezer applications employ very stiff traps (large $\kappa$) to ensure that a trapped particle does not hop out of a given potential well. 
However, in our experiments, we tune the laser power (vary $\kappa$) to explore the effect of trapping strength on the dispersion of particles.
To study dispersion in dynamic potential fields, we oscillated all traps synchronously with the sinusoidal velocity,
\begin{equation}
    \bm{u}(t) = \hat{\bm{e}}_{x} \omega A \cos{(\omega t)},
    \label{eq:trap_velocity}
\end{equation}
where $A$ is the amplitude and $\omega$ is the angular frequency.
Upon depositing a dilute concentration of silica beads with radius $a=1.25$ $\mu$m to the bottom of an imaging chamber, we observed oscillatory motion as the particles moved in-and-out of neighboring harmonic wells along the 2D plane.
We tracked the particle trajectories and measured their long-time self diffusivity using optical microscopy. Further details on our experimental methodology can be found in the Supplemental Material \cite{Supplemental}.

%
%

We apply generalized Taylor dispersion theory \cite{Brenner1993} to understand the coupling between oscillatory trap motion and colloidal diffusion.
For a Brownian particle that enters an $L\times L$ cell occupied by a moving harmonic trap, the normalized probability density $g(\bm{r}, t)$ of finding the particle at a position $\bm{r}$ and time $t$ is governed by the Smoluchowski equation,
\begin{equation}
	\left(
	\frac{\partial}{\partial t} 
	+ 
	\mathscr{L} 
	\right)
	g(\bm{r}, t)
	=
	0
	,
	\label{eq:g_eqn}
\end{equation}
where
\begin{equation}
	\mathscr{L}(\,\cdot\,)
	=
	\bm{u} (t) \cdot \bm{\nabla}_{\bm{r}} (\,\cdot\,)
	-
	\frac{kT}{\gamma}
	\nabla_{\bm{r}}^{2} (\,\cdot\,)
	-
	\frac{1}{\gamma}
	\bm{\nabla}_{\bm{r}} \cdot[(\,\cdot\,) \bm{\nabla}_{\bm{r}} V (\bm{r})]
	,
	\label{eq:flux}
\end{equation}
is the time-evolution operator, $V(\bm{r})$ is the potential-energy field given by Eq.~\eqref{eq:harmonic_potential}, $\bm{u}(t)$ is the velocity of the moving traps given by Eq.~\eqref{eq:trap_velocity}, $kT = 4.046\times 10^{-21}$ J is the thermal energy, and $\gamma$ is the particle resistivity. The terms on the right-hand side of Eq.~\eqref{eq:flux} reflect transport by convection, diffusion, and potential-energy gradients. The ratio $D_{0} \equiv kT/\gamma$ defines the Stokes-Einstein-Sutherland diffusivity, which we measure to be $D_{0} \approx 0.105$ $\mu$m$^2$/s in the experiments.

Particle density fluctuations give rise to an {\it effective} diffusivity that is distinct from the Stokes-Einstein-Sutherland value. The strength and orientation of these fluctuations are captured by the probability-weighted displacement field $\bm{d}(\bm{r}, t)$, which satisfies the inhomogeneous equation,
\begin{equation}
	\left( \frac{\partial}{\partial t} + \mathscr{L} \right)\bm{d}(\bm{r}, t)
	=
	\frac{2kT}{\gamma} 
	\bm{\nabla}_{\bm{r}} g 
	+
	\frac{1}{\gamma}
	[g \bm{\nabla}_{\bm{r}} V
	-
	\braket{g \bm{\nabla}_{\bm{r}} V}g
	]
	,
	\label{eq:d_eqn}
\end{equation}
where $\braket{\,\cdot\,} \equiv L^{-2}\int_{L^{2}} (\,\cdot\,) \, \mathrm{d} \bm{r}$ denotes the spatial average over an $L \times L$ cell. Clearly, the evolution of $\bm{d}$ is one-way coupled to the evolution of $g$ through the terms on the right-hand side of Eq.~\eqref{eq:d_eqn}. These terms reflect fluctuations in the probability current, which drive long-wavelength disturbances to the number density of particles. Following Brady and coworkers \cite{morris1996self,Zia2010,Takatori2014,burkholder2017tracer,burkholder2019fluctuation,peng2020upstream}, it can be shown that the structure field $g(\bm{r},t)$ is directly related to the effective drift velocity of the particle,
\begin{equation}
	\bm{U} (t)
	=
	\bm{u} (t) 
	-
	\frac{1}{\gamma} \braket{g \bm{\nabla}_{\bm{r}}V} (t)
	,
	\label{eq:effective_drift}
\end{equation}
while the displacement field $\bm{d}(\bm{r},t)$ is related to the effective diffusivity tensor,
\begin{equation}
	\tens{D} (t)
	=
	\frac{kT}{\gamma} \tens{I}
	+
	\frac{1}{\gamma}\braket{\bm{d} \bm{\nabla}_{\bm{r}} V} (t)
	.
	\label{eq:effective_diffusivity}
\end{equation}
The last two expressions are the key results of the dispersion theory. They show that the enhancement (or reduction) in drift and diffusion is driven by the average particle flux down potential-energy gradients. 

Eqs.~\eqref{eq:g_eqn} and \eqref{eq:d_eqn} were solved numerically in an $L \times L$ cell subject to periodic boundary conditions and the normalization conditions $\braket{g} = 1$ and $\braket{\bm{d}} = \bm{0}$. Our numerical solutions were developed using the finite-element method with implicit time-advancement in COMSOL Multiphysics$^\text{\textregistered}$. The resulting $g$- and $\bm{d}$-fields were then inserted into Eqs.~\eqref{eq:effective_drift}-\eqref{eq:effective_diffusivity} to compute the effective drift and diffusivity of the particle as a function of time. We validated the dispersion theory by developing Brownian dynamics simulations of 10,000 freely draining (i.e., non-interacting) particles in HOOMD-blue \cite{anderson2020hoomd} and calculated the diffusivity from the long-time growth of their mean-squared displacements. Further details on the derivation of the relevant equations, numerical method, and simulations can be found in the Supplemental Material \cite{Supplemental}. Below, we present the key results from the theoretical calculations and compare them against the experimental measurements.

%
%

\begin{figure} [!b]
	\vspace{-10pt}
	\centering
	\includegraphics[width=1.0\linewidth]{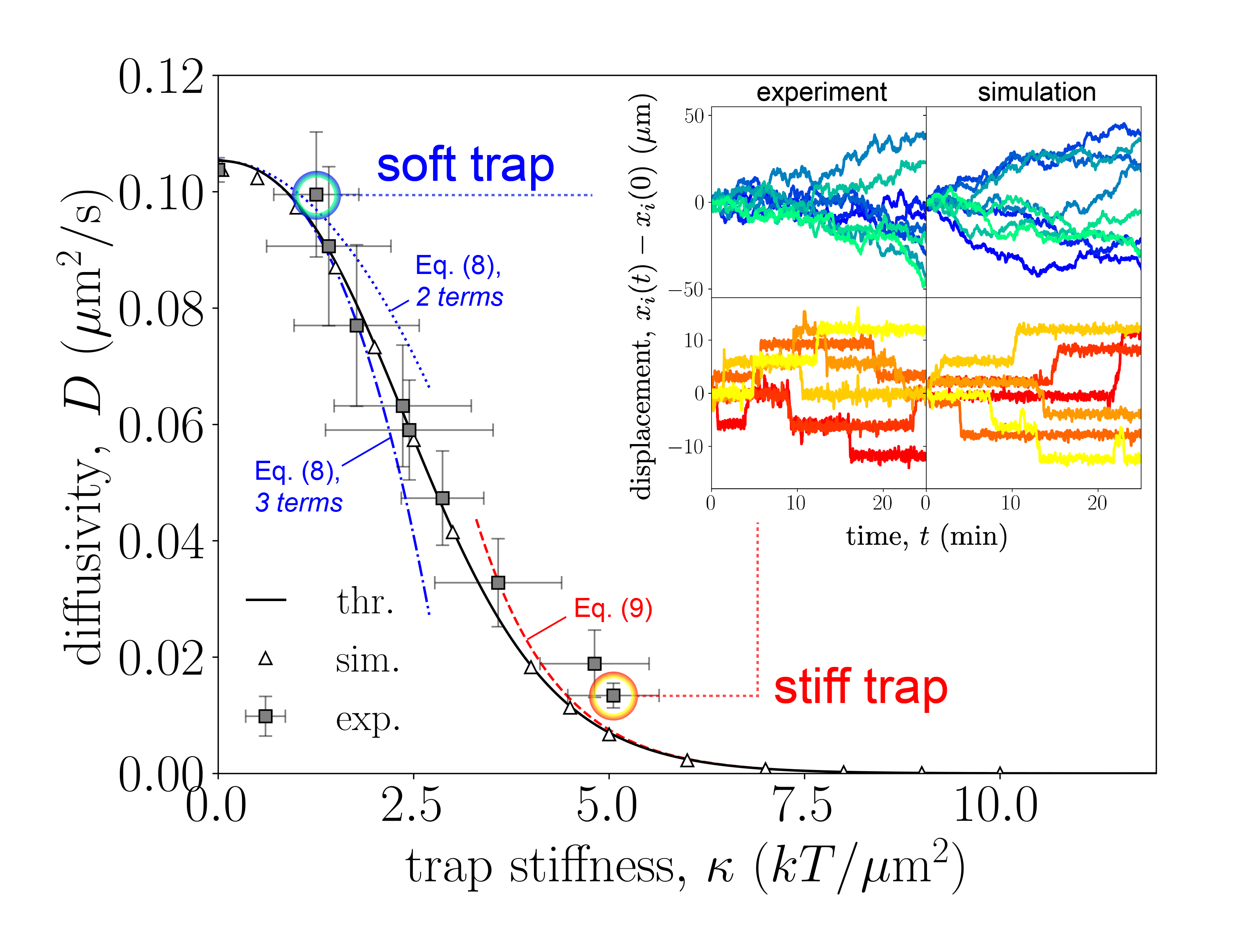}
	\vspace{-10pt}
	\caption{%
		Effective diffusivity $D$ of particles in stationary traps decreases monotonically with trap stiffness $\kappa$. 
		Shown are results from experiments (squares), Brownian dynamics simulations (triangles), Smoluchowski theory (solid line), and asymptotic limits [dashed lines, see Eqs.~\eqref{eq:diffusivity_soft_traps}-\eqref{eq:diffusivity_stiff_traps}]. A proportionality constant of 1.5 was used in Eq.~\eqref{eq:diffusivity_stiff_traps} to fit the numerical data.
		Inset: Particle trajectories from the experiments and simulations indicate random walks for soft traps and activated, Kramers-like hopping for stiff traps (see also Supplemental Movies S1-S2 \cite{Supplemental}).
	}
	\label{fig:Fig2}
\end{figure}

When the traps are held stationary, the convective term in Eq.~\eqref{eq:flux} vanishes and the particle probability distribution achieves a steady state. The absence of a time-dependent convective term in the Smoluchowski equation implies zero net drift, $\bm{U} = \bm{0}$, and an isotropic, time-independent diffusivity $\tens{D}$ with components $D_{xx} = D_{yy} = D$. Figure \ref{fig:Fig2} shows that the scalar diffusivity $D$ decreases monotonically with the trap stiffness $\kappa$, as reported in previous studies using one-dimensional (1D) potentials \cite{Fredrickson1991}. 
(Supplemental Movies S1-S2 \cite{Supplemental} show measured and simulated particle motion in stationary traps of varying stiffness.)
For ``soft'' traps (i.e., potential well depths $\Delta V \ll kT$), a regular perturbation analysis admits the following expansion for the diffusivity:
\begin{align}
	D
	&=
	\frac{kT}{\gamma} 
	\,
	\bigg(
		1
		-
		\frac{\braket{(V - \braket{V})^{2}}}{2(kT)^{2}}
	\nonumber \\
	& \qquad\qquad
		+
		\frac{\braket{(V - \braket{V})^{3}}+\braket{\bm{\nabla}_{\bm{r}} (|\bm{\nabla}_{\bm{r}}\varPhi|^{2}) \cdot \bm{\nabla}_{\bm{r}} V}}{4 (kT)^{3}}
	\nonumber \\
	& \qquad\qquad
		+
		\cdots
	\bigg)
	,
	\label{eq:diffusivity_soft_traps}
\end{align}
where $\varPhi (\bm{r})$ satisfies $\nabla^{2}_{\bm{r}} \varPhi (\bm{r}) = \braket{V} - V(\bm{r})$ and $\braket{\varPhi} = 0$. Equation \eqref{eq:diffusivity_soft_traps} indicates that the reduction in diffusivity below the Stokes-Einstein-Sutherland value is proportional to the spatial variance in the potential energy; both the first and second corrections are plotted in Fig.~\ref{fig:Fig2}. In this regime, the particle trajectories appear to follow a random walk as in classical Brownian motion (see Fig.~\ref{fig:Fig2}, upper panel of inset).

For ``stiff'' traps ($\Delta V \gg kT$) held in a fixed configuration, the particles undergo activated-hopping dynamics and their diffusivity is very nearly zero. Any given particle remains trapped in a local potential well for a long time, punctuated by discrete transitions (``hops'') from one well to another (see Fig.~\ref{fig:Fig2}, lower panel of inset). Kramers' theory \cite{kramers1940brownian,brinkman1956brownian,brinkman1956brownian2} suggests that the effective diffusivity is proportional to the characteristic ``hopping frequency,'' which scales linearly with the curvature of the potential well $\kappa = \tfrac{1}{2}(\nabla_{\bm{r}}^{2} V)|_{\bm{r}=\bm{0}}$ and exponentially with the well depth $\Delta V = \tfrac{1}{8} \kappa W_{\text{trap}}^{2}$:
\begin{equation}
	D
	\propto
	\frac{L^{2}}{4 \uppi \gamma}
	\mathrm{e}^{-\Delta V/kT}
	(\nabla_{\bm{r}}^{2} V)|_{\bm{r}=\bm{0}}
	.
	\label{eq:diffusivity_stiff_traps}
\end{equation}
The last relationship is not exact. A constant of proportionality, which would convert Eq.~\eqref{eq:diffusivity_stiff_traps} into an equality, depends upon the ratio $W_{\text{trap}}/L$ between the size and spacing of the harmonic traps. For traps of diameter $W_{\text{trap}} = 3.2$ $\mu$m spaced a distance $L=6$ $\mu$m apart, a proportionality constant of 1.5 gives quantitative agreement with the exact dispersion theory (see Fig.~\ref{fig:Fig2}).
[See the Supplemental Material \cite{Supplemental} for the derivation of Eqs.~\eqref{eq:diffusivity_soft_traps} and \eqref{eq:diffusivity_stiff_traps}.]

The situation qualitatively changes when the traps are not stationary, but oscillated synchronously with the velocity prescribed by Eq.~\eqref{eq:trap_velocity}. After a sufficiently long time, the system achieves a periodic steady state; one is then only interested in time-averaged quantities over a periodic cycle, $\overline{(\,\cdot\,)} \equiv \lim_{\tau \rightarrow \infty} (2\uppi / \omega)^{-1} \int_{\tau-\uppi/\omega}^{\tau+\uppi /\omega} (\,\cdot\,) \, \mathrm{d} t$. It is straightforward to show that the time-averaged drift is identically zero, $\overline{\bm{U}} = \bm{0}$, whereas the time-averaged diffusivity $\overline{\tens{D}}$ is generally non-zero and {\it anisotropic} ($\overline{D}_{xx} \not= \overline{D}_{yy}$) due to the existence of a preferred direction along the convection ($x$-)axis. 

\begin{figure}[!t]
	\begin{center}
		\includegraphics[width=1\linewidth]{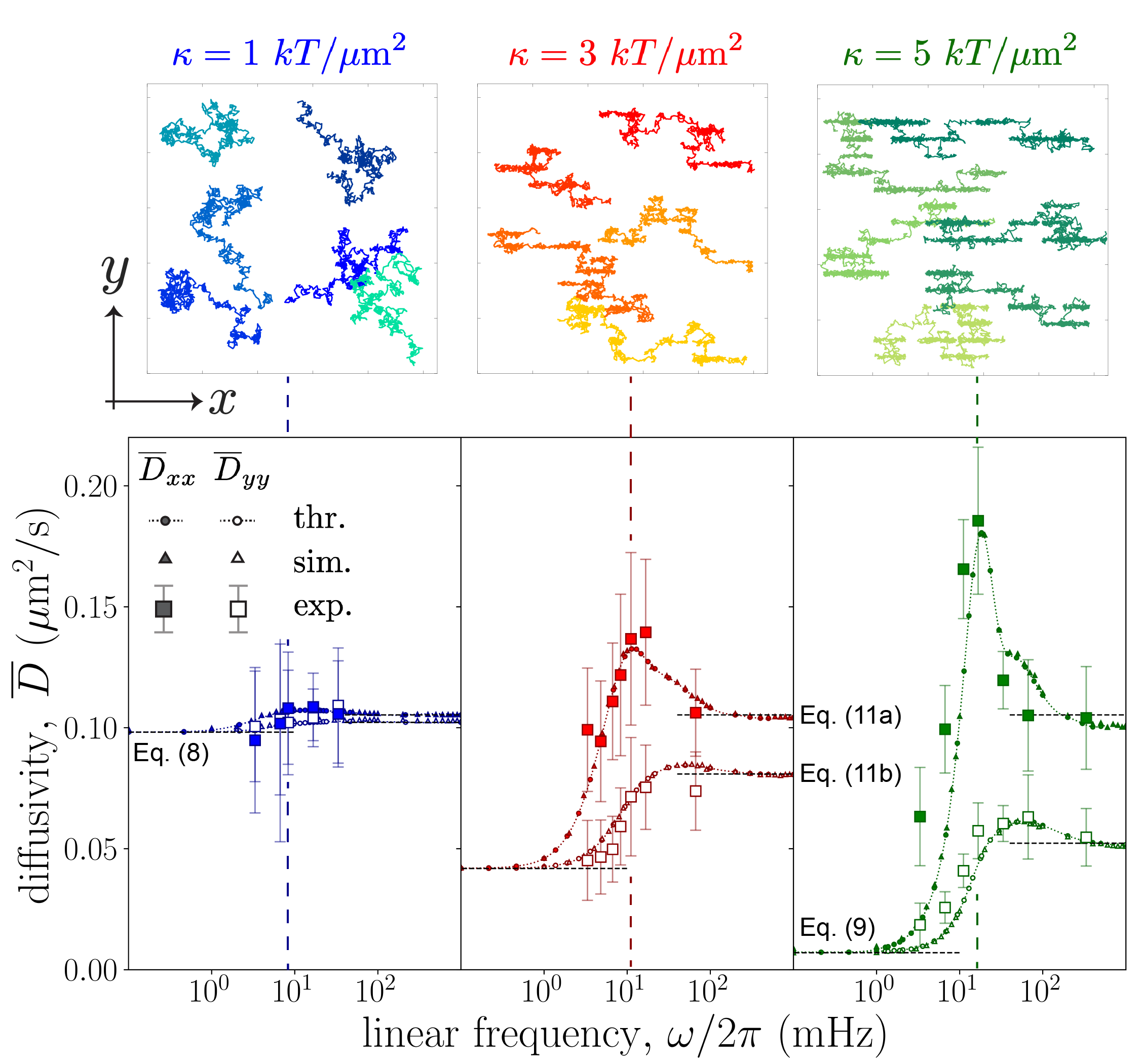}
		\caption{%
			 Oscillating array of harmonic traps generates a non-monotonic, anisotropic dispersion of Brownian particles.
			 (bottom) Time-averaged effective diffusivities $\overline{D}_{xx}$ (filled symbols) and $\overline{D}_{yy}$ (open symbols) plotted as a function of oscillation frequency $\omega$ for different trap stiffnesses $\kappa$. Shown are results from experiments (squares), Brownian dynamics simulations (small triangles), Smoluchowski theory (small circles), and asymptotic limits [dashed lines, see Fig.~\ref{fig:Fig2} and Eqs.~\eqref{eq:diffusivity_soft_traps}, \eqref{eq:diffusivity_stiff_traps}, and \eqref{eq:diffusivity_high_frequency}]. There are no fitting parameters in the theory.
			 (top) Experimental particle trajectories at the critical frequency $\omega_{\text{max}}$, where $\overline{D}_{xx} = \overline{D}_{xx,\text{max}}$, depict increasingly anisotropic dispersion as the trap stiffness is increased.
			 The field of view is $100~\mu$m $\times$ $100~\mu$m.
			 See also Supplemental Movies S3-S4 \cite{Supplemental} for measured and simulated particle trajectories.
		}
		\label{fig:Fig3}
	\end{center}
	\vspace{-18pt}
\end{figure}

Figure \ref{fig:Fig3} illustrates the non-monotonic dependence of the time-averaged diffusivities $\overline{D}_{xx}$ and $\overline{D}_{yy}$ with the driving frequency $\omega$ for three different trap stiffnesses $\kappa = 1$, $3$, and $5$ $kT$/$\mu$m$^2$ and a fixed amplitude $A=5$ $\mu$m. The softest of these traps ($\kappa = 1$ $kT$/$\mu$m$^2$) exhibits the weakest coupling between convection and potential-energy gradients: over a broad range of frequencies, diffusion remains nearly isotropic and close to the Stokes-Einstein-Sutherland limit $D_{0} \approx 0.105$ $\mu$m$^2$/s. As the trap stiffness is increased to $\kappa = 3$ and 5 $kT$/$\mu$m$^2$, the diffusivity becomes increasingly anisotropic with faster diffusion in the oscillating direction relative to the transverse direction ($\overline{D}_{xx} > \overline{D}_{yy}$). Tracking the particle trajectories, depicted at the top of Fig.~\ref{fig:Fig3}, visually confirms the anisotropic dispersion (Supplemental Movies S3-S4 \cite{Supplemental} show measured and simulated trajectories in oscillating traps of varying frequency and fixed stiffness).
Both $\overline{D}_{xx}$ and $\overline{D}_{yy}$ increase to a maximum before decaying to an asymptotic plateau as $\omega$ becomes infinitely large (``ultrafast cycling''). Varying the oscillation amplitude $A$ at fixed frequency $\omega$ reveals a similar, non-monotonic trend (additional data provided in the Supplemental Material \cite{Supplemental}).

The high-frequency asymptote can be understood as follows. Over a time increment much shorter than the Brownian time, a particle samples the entire potential range along the convection axis as the potential field is rapidly cycled. 
Therefore, the effective potential that is ``felt'' by the particle over one periodic cycle is approximated by averaging $V$ over the convection axis:
\begin{equation}
	v(y) = \frac{1}{L} \int_{-L/2}^{L/2} V(x,y) \, \mathrm{d} x.
	\label{eq:1d_potential}
\end{equation}
The quasi-steady diffusion of a Brownian particle in a 1D potential $v(y)$ is well established \cite{lifson1962self,Festa1978}, with diffusivities (derived in the Supplemental Material \cite{Supplemental}),
\begin{subequations}
\begin{flalign}
	\overline{D}_{xx} 
	&= \frac{kT}{\gamma},
	\\
	\overline{D}_{yy}
	&=
	\frac{kT}{\gamma}
	\braket{\mathrm{e}^{-v/kT}}^{-1}
	\braket{\mathrm{e}^{v/kT}}^{-1}
	.
\end{flalign}
	\label{eq:diffusivity_high_frequency}%
\end{subequations}
Equation \eqref{eq:diffusivity_high_frequency} agrees well with the data plotted in Fig.~\ref{fig:Fig3} at the highest of frequencies. Whereas diffusion perpendicular to convection is hindered as though the particle experienced a potential-energy field given by Eq.~\eqref{eq:1d_potential}, parallel diffusion is largely unaffected because the potential-energy gradients along the $x$-direction have essentially been ``smeared out.'' Put another way: since the time required for a Brownian particle to diffuse from one lattice site to another is much slower than the convection time ($\gamma L^{2} / kT \gg 2\uppi/\omega$), the particle is unable to quickly respond to the rapid motion of the traps as it freely diffuses along the convection axis.

\begin{figure}[!t]
	\begin{center}
		\includegraphics[width=1\linewidth]{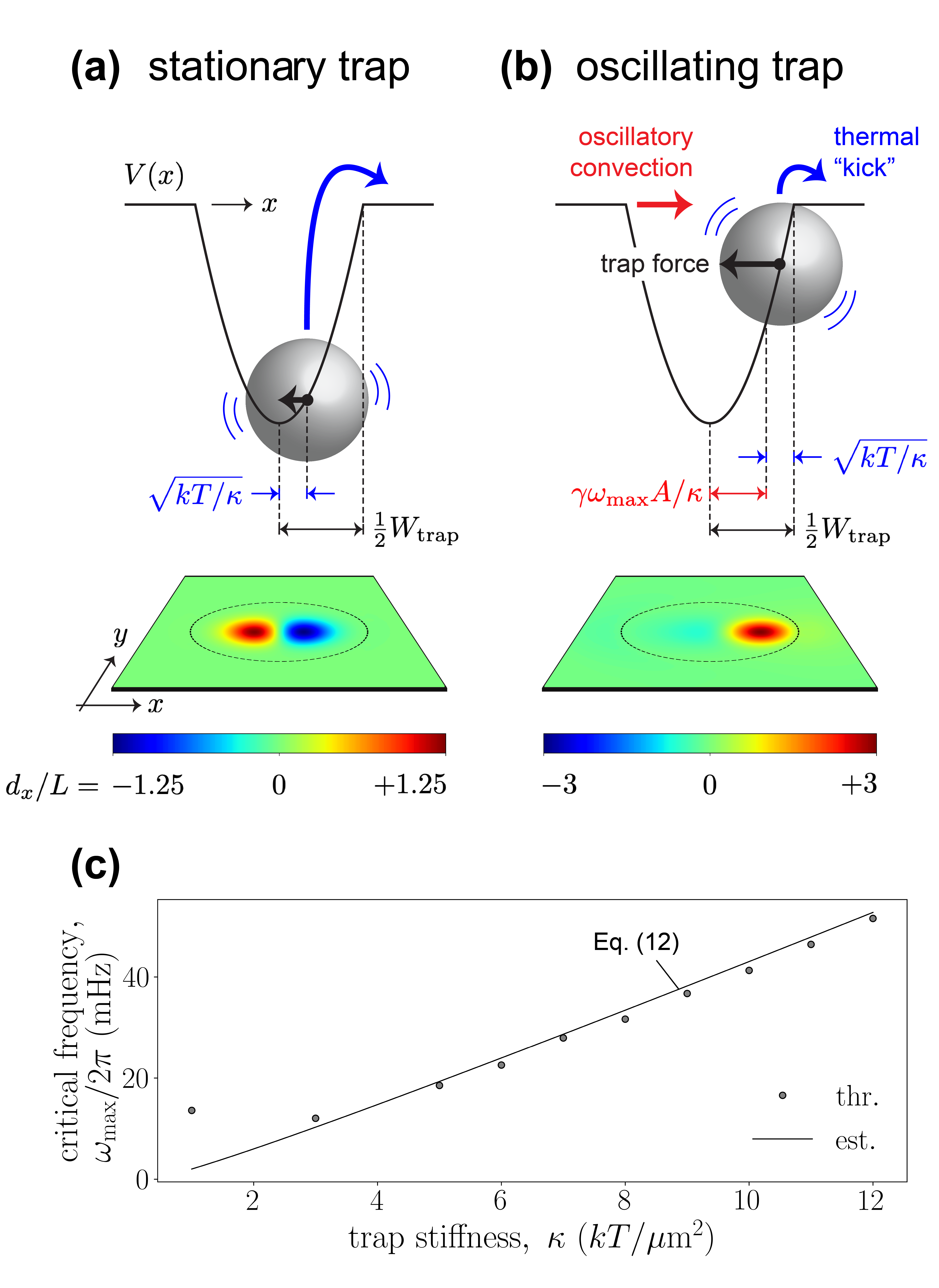}
		\caption{%
			``Slingshot'' mechanism of enhanced dispersion in an oscillating array of harmonic traps. (a) A particle trapped in a stationary potential-energy well undergoes $O(\sqrt{kT/\kappa})$ positional fluctuations due to Brownian motion. Iso-contours of the displacement field density $d_{x}$ reveal a dipolar profile. (b) Oscillation at the critical frequency $\omega_{\text{max}}$ convects the particle probability up the potential-energy gradient by an $O(\gamma \omega_{\text{max}} A/\kappa)$ distance, effectively lowering the barrier to escape. The convected $d_{x}$-field samples larger trapping forces, resulting in enhanced dispersion along the convection axis. Contour plots in (a,b) were generated for $\kappa = 5$ $kT$/$\mu$m$^2$. See also Supplemental Movies S5-S7 for simulated particle trajectories and displacement field densities. (c) The critical frequency $\omega_{\text{max}}$ plotted as a function of the trap stiffness $\kappa$ favorably agrees with the rough estimate given by Eq.~\eqref{eq:critical_frequency}.
		}
		\label{fig:Fig4}
	\end{center}
	\vspace{-15pt}
\end{figure}

Surprisingly, both theory and experiment predict a diffusivity maximum that exceeds the Stokes-Einstein-Sutherland value, $\overline{D}_{xx,\text{max}} > D_{0}$, at a critical oscillation frequency $\omega_{\text{max}}$ (see Fig.~\ref{fig:Fig3}). Figure \ref{fig:Fig4}a,b sketches the basic argument for this maximum. In a stationary system, a strongly trapped Brownian particle fluctuates with variance $kT/\kappa$ about a local potential-energy minimum until a sufficiently large, thermal ``kick'' successfully propels the particle out of the potential well and into the interstices of the lattice (see Fig.~\ref{fig:Fig4}a, top and Supplemental Movie S5 \cite{Supplemental}). Oscillatory convection displaces the particle along the $x$-axis with amplitude $A[1+(\kappa/\gamma \omega)^{2}]^{-1/2} \approx \gamma \omega A / \kappa$, bringing it towards the edge of the trap at $x = \pm \tfrac{1}{2} W_{\text{trap}}$ and effectively lowering the barrier to escape (see Fig.~\ref{fig:Fig4}b, top and Supplemental Movie S6 \cite{Supplemental}). Consequently, the particle is never trapped for very long, but rather is catapulted between lattice sites through the motion of the harmonic traps. This ``slingshot'' mechanism is facilitated at a critical frequency $\omega_{\text{max}}$ for which the fluctuating particle position (with mean $\sim \gamma \omega_{\text{max}} A / \kappa$ and variance $\sim kT/\kappa$) is convected a distance $\tfrac{1}{2} W_{\text{trap}}$ up the potential-energy gradient. By this argument, we make the following estimate for $\omega_{\text{max}}$ (derived in the Supplemental Material \cite{Supplemental}):
\begin{equation}
	\omega_{\text{max}}
	\approx
	\frac{\kappa}{\gamma A} \left( \tfrac{1}{2} W_{\text{trap}} - \sqrt{\frac{kT}{\kappa}} \right)
	.
	\label{eq:critical_frequency}
\end{equation}
This rough estimate qualitatively predicts the critical frequency $\omega_{\text{max}}$ over a range of trap stiffnesses $\kappa$ and quantitatively up to a relative error of about 5\% above the exact calculation (Fig.~\ref{fig:Fig4}c).

The enhanced dispersion can also be rationalized by plotting the two-dimensional iso-contours of the displacement field density $d_{x}$ with and without convection (see Fig.~\ref{fig:Fig4}a,b, bottom and Supplemental Movie S7 \cite{Supplemental}). Under quiescent conditions, the $d_{x}$-field is strongly localized to the center of the potential well and admits a dipolar profile. Oscillation convects the $d_{x}$-field to the edge of the trap, where the potential-energy gradient $\partial V/\partial x$ is maximized. Larger trapping forces are, therefore, weighted more heavily in the force-displacement dyad $\braket{d_{x} (\partial V / \partial x)}$ that appears in the $xx$-component of Eq.~\eqref{eq:effective_diffusivity}. 
This argument directly explains the diffusivity maximum $\overline{D}_{xx,\text{max}}$ observed at the critical frequency $\omega_{\text{max}}$.

The fact that dispersion along the convection axis increases significantly with increasing trap stiffness may be counter-intuitive, given that strong harmonic traps reduce the particle diffusivity under quiescent conditions. A useful analogy is the classical Taylor-Aris dispersion of a tracer in a pressure-driven fluid flow \cite{Aris1956,Taylor1953}, in which {\it smaller} tracer diffusivities generate stronger dispersion along the convection axis due to the coupling between longitudinal convection and transverse diffusion. This effect becomes more pronounced with increasing convection strength. In our system, the strongest dispersion occurs when convection, diffusion, and potential-energy gradients are all in play and on equal footing. If the traps are too stiff, then the particles remain confined to their wells at the mercy of thermal forces; too strong a convective velocity, and the particles are swept past the wells and only sense transverse gradients in the potential-energy landscape. The ``optimal'' rate of convection, for a given trap stiffness, oscillation amplitude, and particle size, is satisfactorily predicted by Eq.~\eqref{eq:critical_frequency}.

%
%

We end this Letter by providing several areas for future investigation.
First, one can easily adapt our experimental system to generate other forms of time-dependent trap motion.
This study focused on 1D synchronous, sinusoidal motion for simplicity; asynchronous or anharmonic kinematics will likely give rise to different couplings with the potential-energy field produced by the traps. This, in turn, could either enhance or hinder dispersion and merits further study.
Second, in addition to changing the convective forcing, one could investigate colloids with different packing densities and surface chemistries to understand how dynamic external fields impact multibody interactions (including hydrodynamic interactions) and macroscopic suspension properties.
Finally, the use of self-propelled colloids would generate further couplings with the dynamic potential landscape, producing nontrivial effects that could be relevant to the field of active matter.

%
%

This material is based upon work supported by the National Science Foundation (Grant No. 2150686).
J.M.B. acknowledges support from the National Institute of Health F32 Ruth L. Kirschstein National Research Service Award (Grant No. F32HL156366).

%
%

%

%
%
\clearpage 
\onecolumngrid

\begin{center}
	\textbf{\large Enhanced dispersion in an oscillating array of harmonic traps} \\[7pt]
	\textbf{\large Supplemental Material} \\[16pt]
	Joseph M.\ Barakat and Sho C.\ Takatori
\end{center}

\tableofcontents

\section{1. Experimental Methodology} \label{sec:experimental_methodology}

\subsection{Preparation of lipid-coated particles}

Fluorescently labeled, lipid-coated particles were created by coating silica micro-beads with a supported lipid bilayer (SLB) containing a minority fraction of fluorescently tagged lipid. 
1,2-dioleoyl-sn-glycero-3-phos-phocholine (DOPC) and 1,2-dioleoyl-sn-glycero-3-phospho-L-serine (DOPS) were purchased from Avanti Polar Lipids.
Atto 647-1,2-dioleoyl-sn-glycero-3-phosphoethanolamine (DOPE-Atto 647) was purchased from ATTO-TEC GmbH.
Silica microspheres (diameter 2.5 $\mu$m; catalog code: SS05000) were purchased from Bangs Laboratories.
Small unilamellar vesicles (SUVs) were formed using an established sonication method \cite{bakalar2018size}.
In brief, a lipid film containing DOPC, 5\% DOPS, and 0.5\% DOPE-Atto 647 was dried under nitrogen and then under vacuum for 30 minutes.
The film was rehydrated in Milli-Q (MQ) water to 0.2 mg/mL lipids, sonicated at low power using a tip sonicator (Branson SFX250 Sonifier) at 20\% of maximum, 1 s/2 s on/off, for three minutes. 
MOPS buffer was added at a final concentration of 50 mM MOPS, pH 7.4, 100 mM NaCl to the resulting SUV mixture.

Silica microspheres were cleaned using a 3:2 mixture of sulfuric acid:hydrogen peroxide (Piranha) for 30 minutes in a bath sonicator, spun at 1000 g, and washed 3 times before being resuspended in MQ water. 
To form SLBs on the beads, 50 $\mu$L of SUV solution was mixed with 10 $\mu$L of the cleaned bead suspension. 
The bead/SUV mixture was incubated for 15 minutes at room temperature while allowing the beads to sediment to the bottom of the centrifuge tube. 
Beads were washed 5 times with MQ water by gently adding/removing the liquid without resuspending the beads into solution. 
The fluidity of the SLB was verified by imaging beads on a glass coverslip at high laser intensity, where the diffusion of labeled lipids was visible after photo-bleaching a small region. 
Lipid-coated beads were deposited into a chamber containing MQ water and sealed off to eliminate drift.
The beads settled down to the bottom of the chamber and all experiments were conducted in 2D.

\subsection{Optical tweezer setup and calibration}

An array of moving harmonic traps was generated using optical tweezers (Tweez 305, Aresis Ltd; Ljubljana, Slovenia), using an IR laser (1064 nm) with a maximum power of 5 W continuous wave (CW).
We selected a trap-to-trap switching rate of 100 kHz to ensure that the particles will effectively feel a continuous harmonic potential.
We used a 16 $\times$ 16 array of traps, which results in $\approx 2.5$ ms time delay to illuminate all trap positions.
This time delay is significantly smaller than the Brownian and oscillatory convection timescales in our system, ensuring that the particles experience a continuous harmonic potential.
A custom MATLAB script was written to construct a time trajectory of oscillatory trap positions for each cell lattice position and incorporated into the tweezer software.
The trap focus was adjusted to the mid-plane of the colloids sitting above the substrate.
Laser powers were adjusted from 0.05-0.5 W to vary the trap stiffness from $\kappa = 0.5$-6 $kT/\mu\mathrm{m}^2$.

\begin{figure}[!h]
	\begin{center}
		\includegraphics[width=0.7\linewidth]{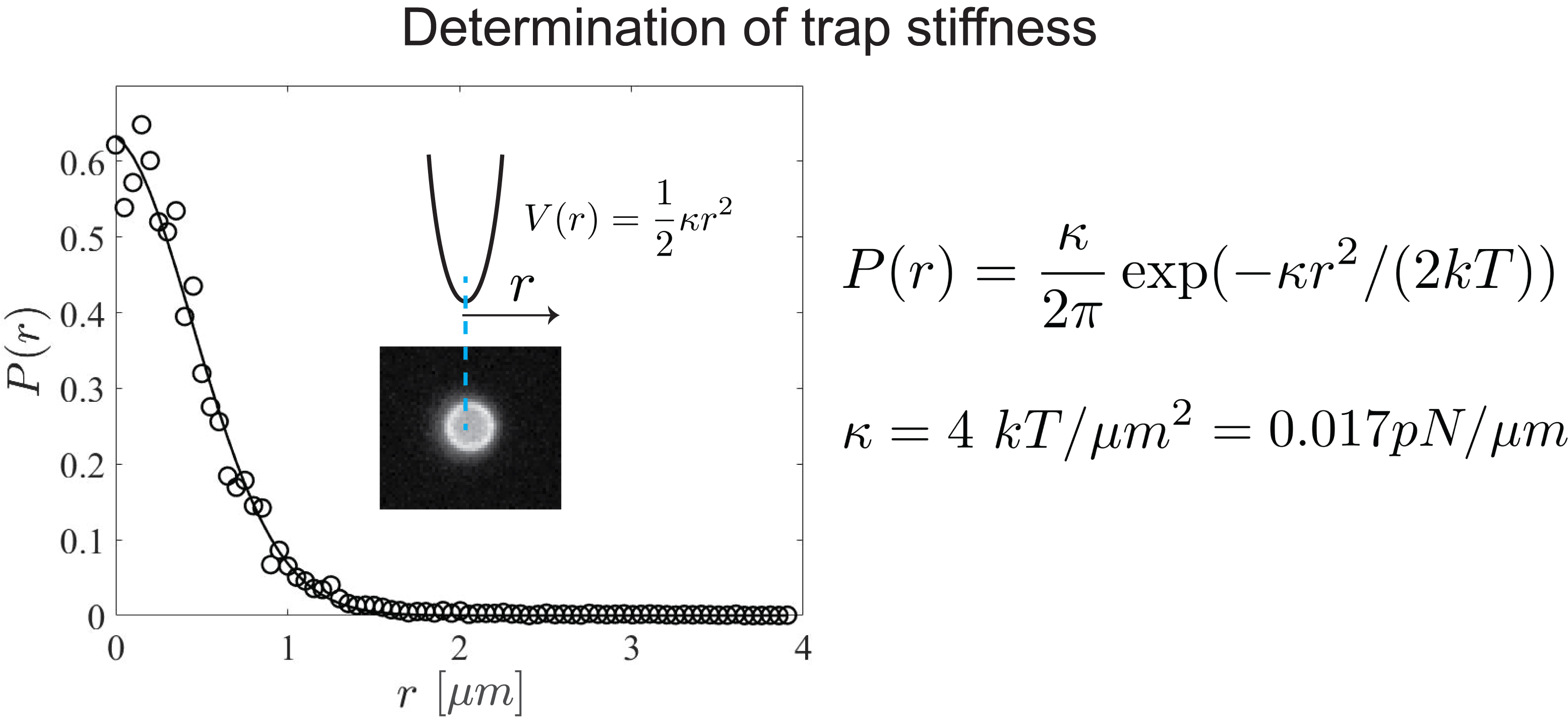}
		\caption{%
			Measurement of trap stiffness $\kappa$ from the equilibrium probability distribution of particles diffusing in a harmonic well generated by optical tweezers.  
			Data are fit to a Boltzmann distribution to obtain $\kappa$ ($\kappa = 4~kT /\mu\mathrm{m}^2$ in the case shown).
			This measurement was averaged over all 16 $\times$ 16 trap positions in the lattice array and repeated for every laser power used in this study.
		}
		\label{Fig:SI1}
	\end{center}
	\vspace{-18pt}
\end{figure}

The trap stiffness $\kappa$ was calibrated by measuring the equilibrium probability distribution of the particles in a stationary array of traps.
For each laser power, $\kappa$ was obtained by binning particles by their radial position $r$ from the center of the trap and fitting the binned data to a Boltzmann distribution, $P(r) = (\kappa/2\pi)\mathrm{e}^{-\kappa r^{2}/(2kT)}$.
An example of a distribution and fit is shown in Fig.~\ref{Fig:SI1}.
We verified that there are no variations in trip stiffness between different lattice positions in the array.

\begin{figure}[!h]
	\begin{center}
		\includegraphics[width=0.7\linewidth]{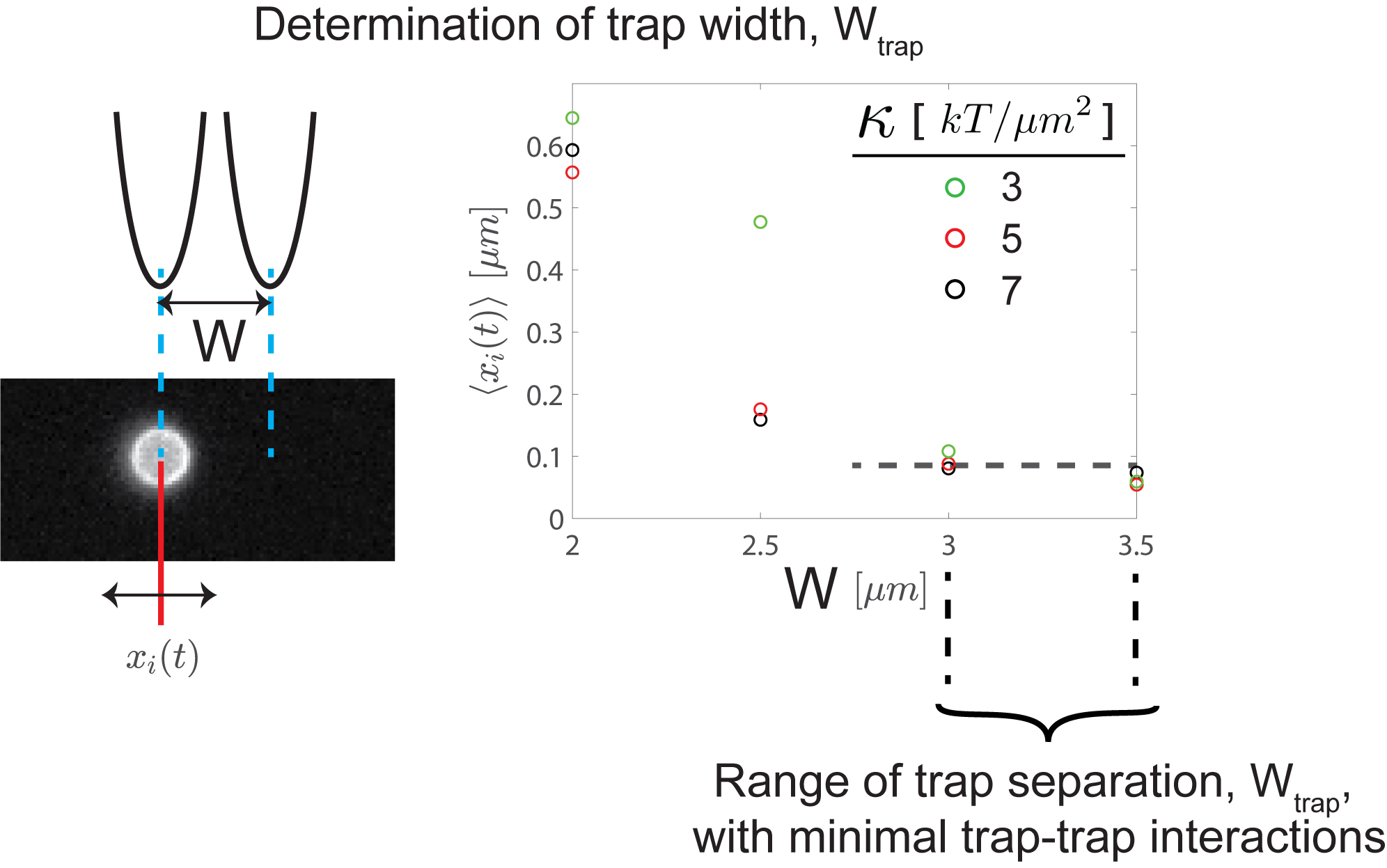}
		\caption{%
			Measurement of trap width $W_{\text{trap}}$.
			A second trap was placed at varying separation distances from the first trap containing a trapped bead. 
			We measured the time-averaged position of the trapped bead, $\langle x_i(t) \rangle$, for varying separation distances at fixed trap stiffness.
			We found that the average position is pulled towards the second trap at distances $W < 3$ $\mu$m and is approximately constant in the range $W = 3$-3.5 $\mu$m. This gives an average trap width $W_{\text{trap}} \approx 3.2$ $\mu$m.
		}
		\label{Fig:SI2}
	\end{center}
	\vspace{-18pt}
\end{figure}

The trap width $W_{\text{trap}}$ was determined from a separate set of experiments.
Two traps were placed side-by-side with center-to-center separation distance $W$.
The first trap, containing a trapped particle, was held fixed while the position of the second trap was varied; the average position $\langle x_{i}(t) \rangle$ of the particle was measured as a function of the separation distance $W$ (Fig.~\ref{Fig:SI2}).
When the second trap is placed far away, no interference is observed on the average position of the particle.
However, as the second trap is moved closer, $W < 3$ $\mu$m for a particle of radius $a = 1.25$ $\mu$m, the average position drifts towards the second trap.
We found that the average particle position remains approximately constant within the range of separation distances of $W =$ 3-3.5 $\mu$m, giving an approximate trap width $W_{\text{trap}} \approx 3.2$ $\mu$m.

\subsection{Measurement of diffusivity}

The long-time self diffusivity was determined by particle tracking. All imaging was carried out on an inverted Nikon Ti2-Eclipse microscope (Nikon Instruments) using a water-immersion objective (Plan Apochromat VC 60x, numerical aperture 1.2, water). 
Lumencor SpectraX Multi-Line LED Light Source was used for excitation (Lumencor, Inc).
Fluorescent light was spectrally filtered with an emission filter (680/42; Semrock, IDEX Health and Science) and imaged on a Photometrics Prime 95 CMOS Camera (Teledyne Photometrics).
In order to achieve satisfactory long-time statistics, particle trajectories were measured for times much larger than all other timescales in the system (including the diffusive timescale $\gamma L^{2}/kT$, oscillation period $2\uppi/\omega$, and trapping timescale $\gamma/\kappa$). A modified MATLAB script, based on the IDL code by Crocker and Grier \cite{crocker1996methods,crockerweeksIDL,blairdufresneMATLAB}, was used to track the individual particles by identifying each particle center and tracking its trajectory over time using an image stack with one frame taken every 1-2 s. 
Particles that were immobile (due to defects) were filtered out so as not to be considered during image post-processing.

\begin{figure}[!h]
	\begin{center}
		\includegraphics[width=0.5\linewidth]{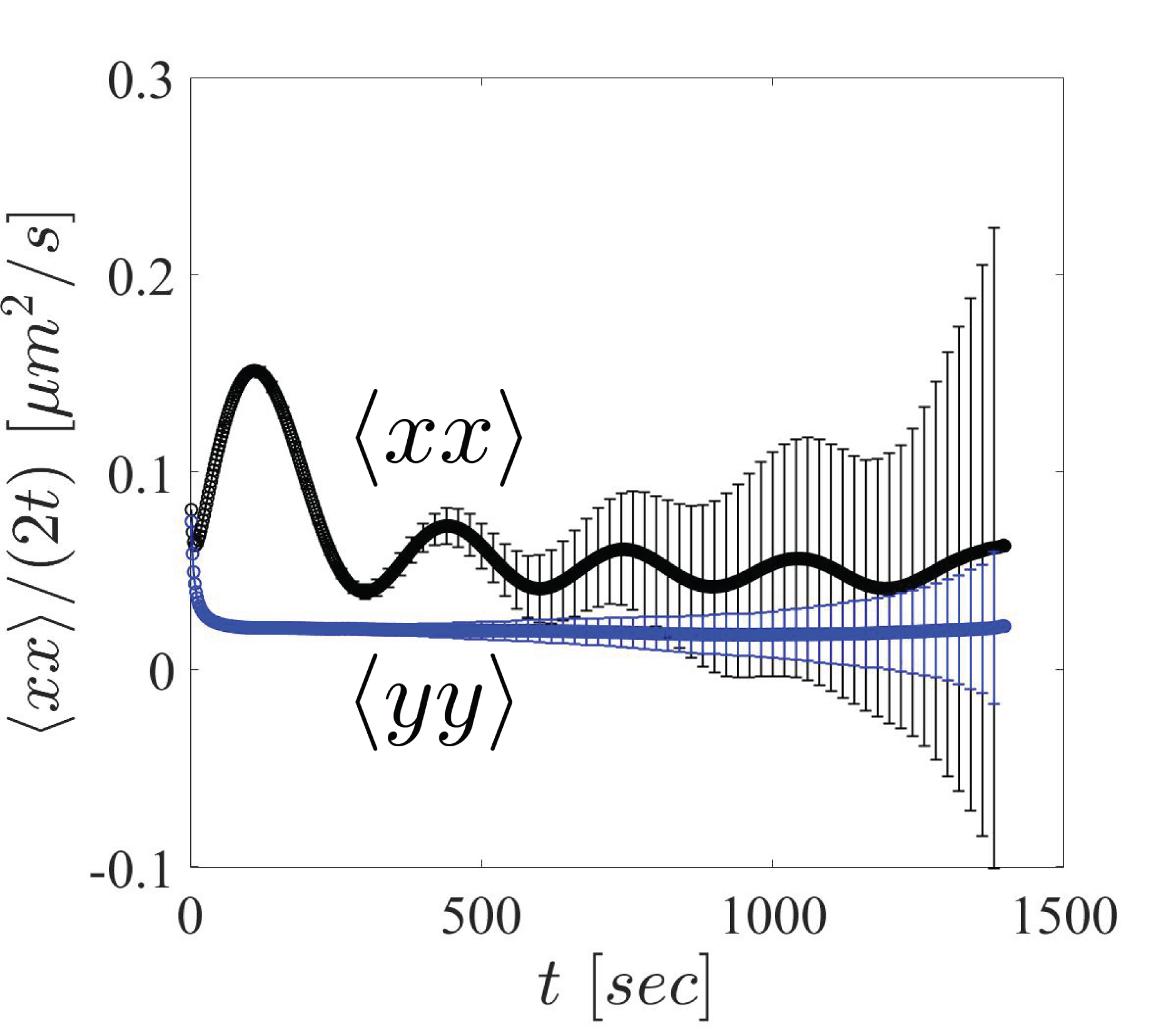}
		\caption{%
			Representative mean squared displacements $\langle \Delta x(t) \Delta x(t) \rangle / (2t)$ (black symbols) and $\langle \Delta y(t) \Delta y(t) \rangle / (2t)$ (blue symbols) measured using Eq.~\eqref{eq:SI_msd_experiments} for Brownian particles diffusing through an oscillating array of harmonic traps.
			Diffusivities reported in the main text were computed from the long-time plateaus of these curves, using Eq.~\eqref{eq:SI_diffusivity_experiments}. Statistical errors were calculated using the bootstrap algorithm \cite{ross2020introduction} over the entire observation time window.
		}
		\label{Fig:SI3}
	\end{center}
	\vspace{-18pt}
\end{figure}

The average diffusivity tensor is classically defined in terms of the long-time derivative of the mean squared displacements (MSD) of the particles: 
\begin{equation}
    \overline{\tens{D}} = \lim_{t \rightarrow \infty} \frac{1}{2}
    \frac{\mathrm{d}}{\mathrm{d}t}
    \langle \Delta \bm{R}(t) \Delta \bm{R}(t) \rangle
    ,
    \label{eq:SI_diffusivity_tensor}
\end{equation}
where $\bm{R}$ denotes the {\it global} position vector [related to the {\it local} position vector $\bm{r}$ by Eq. \eqref{eq:SI_coordinate_conversion}, below] and the angle brackets $\langle\,\cdot\,\rangle$ denote an {\it ensemble} average (not to be confused with the {\it cell} average defined in the main text). The MSD tensor over a time interval $t$ is computed from the formula,
\begin{equation}
	\braket{
	\Delta \bm{R} (t)
	\Delta \bm{R} (t)
	}
	=
	\frac{1}{N_{\text{p}}}
	\sum_{i=1}^{N_{\text{p}}}
	\lim_{\tau\rightarrow\infty}
	\frac{1}{\tau-t}
	\int_{0}^{\tau-t} 
	\left[ \bm{R}_{i}(s + t) - \bm{R}_{i}(s)  \right]
	\left[ \bm{R}_{i}(s + t) - \bm{R}_{i}(s)  \right]
	\, \mathrm{d} s
	,
	\label{eq:SI_msd_experiments}
\end{equation}
where $\bm{R}_{i}(t)$ denotes the global position of the $i$th particle at time $t$.
In Eq.~\eqref{eq:SI_msd_experiments}, the squared displacement of a particle with index $i$ is first averaged over all time windows of duration $t$ within the interval $\tau$ of the particle's trajectory. This ``time average'' for each $i$th particle, evaluated in the limit as $\tau \rightarrow \infty$, is subsequently averaged over all particles $i = 1, 2, \dots, N_{\text{p}}$ to approximate the ensemble average of all squared displacements with satisfactory statistics.
At long times, the MSD tensor $\langle \Delta \bm{R}(t) \Delta \bm{R}(t) \rangle$ oscillates with fixed amplitude about a steady, linear growth. Thus, the long-time derivative of the MSD can be measured by simply dividing by time, leading to the relation,
\begin{equation}
    \overline{\tens{D}} = \lim_{t \rightarrow \infty} \frac{1}{2 t}
     \langle \Delta \bm{R}(t) \Delta \bm{R}(t) \rangle
     .
     \label{eq:SI_diffusivity_experiments}
\end{equation}
Equation \eqref{eq:SI_diffusivity_experiments} was used to measure the diffusivity from the measured particle trajectories (see Fig.~\ref{Fig:SI3}). Trajectories were averaged over a sufficiently long time interval $\tau$ to ensure linear growth, and the time integral in Eq.~\eqref{eq:SI_msd_experiments} was discretized using the left Riemann sum.
Statiscal errors in the MSD were calculated using a bootstrap algorithm \cite{ross2020introduction}.

The particle resistivity $\gamma$ used in all theoretical calculations was calibrated by measuring the Stokes-Einstein-Sutherland diffusivity $D_{0} = kT/\gamma \approx 0.105$ $\mu$m$^2$/s of particles diffusing in the absence of a harmonic potential. For a spherical particle of radius $a$ in a fluid of viscosity $\eta$, the particle resistivity is given by $\gamma = 6 \uppi \eta a K_{D}$, where $K_{D}$ is a drag-correction factor to account for the hydrodynamic interaction with a nearby wall (in our case, the substrate floor). For our system with $a = 1.25$ $\mu$m and $\eta = 1$ cP, we estimate the drag-correction factor to be $K_{D} = kT/(6\uppi \eta a D_{0}) \approx 1.63$, corresponding to a particle-to-wall spacing of about 0.5 $\mu$m according to Fax\'en's formula \cite{happel2012low}. This gives a particle resistivity of $\gamma \approx 9.49$ $kT\cdot\text{s}/\mu\text{m}^2$.

\section{2. Taylor-Dispersion Theory} \label{sec:dispersion_theory}

\subsection{Derivation of Eqs. (3)-(5): governing equations for the probability density and displacement}

The starting point for deriving the basic equations in the main text is the single-particle Smoluchowski equation,
\begin{equation}
	\frac{\partial P(\bm{R}, t)}{\partial t}
	=
	-
	\bm{\nabla}_{\bm{R}} \cdot \bm{J} (\bm{R}, t)
	,
	\label{eq:SI_smoluchowski_eqn}
\end{equation}
where $P(\bm{R}, t)$ is the probability density of finding a Brownian particle at a {\it global} position $\bm{R}$ and time $t$ and
\begin{equation}
	\bm{J} (\bm{R}, t)
	=
	\bm{u} (t) P
	-
	\frac{1}{\gamma}
	[
	kT \bm{\nabla}_{\bm{R}} P 
	+
	P \bm{\nabla}_{\bm{R}} V (\bm{R})
	]
	\label{eq:SI_probability_flux}
\end{equation}
is the probability flux. The spatial periodicity of the potential-energy field allows us to convert the ``global'' position $\bm{R}$ to the ``local'' position $\bm{r}$ via the transformation,
\begin{equation}
	\bm{R} = \bm{n} L + \bm{r},
	\label{eq:SI_coordinate_conversion}
\end{equation}
where $\bm{n}$ contains the lattice indices of a given periodic cell. In terms of lattice and local coordinates, $V(\bm{R}) \equiv V(\bm{r})$, $P(\bm{R},t) \equiv P_{\bm{n}} (\bm{r}, t)$, and $\bm{J} (\bm{R}, t) \equiv \bm{J}_{\bm{n}} (\bm{r},t)$. 

In the following, we employ the ``flux-averaging'' approach of Brady and coworkers \cite{morris1996self,Zia2010,Takatori2014,burkholder2017tracer,burkholder2019fluctuation,peng2020upstream}.
First, we define the continuous wavevector $\bm{k}$ and apply the discrete Fourier transform $\hat{(\,\cdot\,)} \equiv \sum_{\bm{n}} (\, \cdot\,) \mathrm{e}^{\mathrm{i} \bm{k} \cdot \bm{n} L}$ to Eqs.~\eqref{eq:SI_smoluchowski_eqn}-\eqref{eq:SI_probability_flux}, obtaining
\begin{equation}
	\frac{\partial \hat{P}(\bm{k}, \bm{r}, t)}{\partial t}
	=
	-
	(\mathrm{i} \bm{k} + \bm{\nabla}_{\bm{r}})  \cdot \hat{\bm{J}} (\bm{k}, \bm{r}, t)
	,
	\label{eq:SI_smoluchowski_eqn_ft}
\end{equation}
\begin{equation}
	\hat{\bm{J}} (\bm{k}, \bm{r}, t)
	=
	\bm{u} (t) \hat{P}
	-
	\frac{1}{\gamma}
	[
	kT (\mathrm{i} \bm{k} + \bm{\nabla}_{\bm{r}}) \hat{P} 
	+
	\hat{P} \bm{\nabla}_{\bm{r}} V (\bm{r})
	]
	.
	\label{eq:SI_probability_flux_ft}
\end{equation}
Next, we spatially average Eqs.~\eqref{eq:SI_smoluchowski_eqn_ft}-\eqref{eq:SI_probability_flux_ft} over one periodic cell according to $\braket{\,\cdot\,} \equiv L^{-2}\int_{L^{2}} (\,\cdot\,) \, \mathrm{d} \bm{r}$, apply the divergence theorem, and invoke periodic boundary conditions to obtain the continuity equation,
\begin{equation}
	\frac{\partial \hat{\rho}(\bm{k}, t)}{\partial t}
	=
	-
	\mathrm{i} \bm{k} \cdot \braket{\hat{\bm{J}}} (\bm{k}, t)
	,
	\label{eq:SI_smoluchowski_eqn_ft_avg}
\end{equation}
\begin{equation}
	\braket{\hat{\bm{J}}} (\bm{k}, t)
	=
	\bm{u} (t) \hat{\rho}
	-
	\frac{1}{\gamma}
	[
	kT \mathrm{i} \bm{k} \hat{\rho} 
	+
	\braket{\hat{P} \bm{\nabla}_{\bm{r}}V}
	]
	,
	\label{eq:SI_probability_flux_ft_avg}
\end{equation}
where $\hat{\rho} (\bm{k}, t) \equiv \braket{\hat{P}} (\bm{k}, t)$ is the Fourier-transformed number density. Eqs.~\eqref{eq:SI_smoluchowski_eqn_ft_avg}-\eqref{eq:SI_probability_flux_ft_avg} represent the macroscopic transport equations for the periodic lattice.

Next, we define the structure function $\hat{G}(\bm{k}, \bm{r}, t)$ as
\begin{equation}
	\hat{P}(\bm{k}, \bm{r}, t) 
	=
	\hat{\rho} (\bm{k}, t)
	\hat{G}(\bm{k}, \bm{r}, t)
	.
	\label{eq:SI_conditional_probability}
\end{equation}
Multiplying Eq.~\eqref{eq:SI_smoluchowski_eqn_ft_avg} by $\hat{G}$, subtracting from Eq.~\eqref{eq:SI_smoluchowski_eqn_ft}, and dividing through by $\hat{\rho}$ then gives
\begin{flalign}
	\frac{\partial \hat{G}(\bm{k}, \bm{r}, t)}{\partial t}
	&=
	-
	\hat{\rho}^{-1} [ 
	\mathrm{i} \bm{k} \cdot ( \hat{\bm{J}} - \braket{\hat{\bm{J}}} \hat{G} )
	+
	\bm{\nabla}_{\bm{r}} \cdot \hat{\bm{J}}
	]
	\nonumber\\
	&=
	- \bm{u} (t) \cdot \bm{\nabla}_{\bm{r}} \hat{G}
	+
	\frac{kT}{\gamma}
	\nabla_{\bm{r}}^{2} \hat{G}
	+
	\frac{1}{\gamma}
	\bm{\nabla}_{\bm{r}} \cdot 
	[
	\hat{G} \bm{\nabla}_{\bm{r}} V (\bm{r})
	]
	+
	\mathrm{i} \bm{k} \cdot 
	\left(
	\frac{2kT}{\gamma} \bm{\nabla}_{\bm{r}} \hat{G} 
	+
	\frac{1}{\gamma}
	[
	\hat{G} \bm{\nabla}_{\bm{r}} V (\bm{r})
	-
	\braket{\hat{G} \bm{\nabla}_{\bm{r}}V} \hat{G} 
	]
	\right)
	,
	\label{eq:SI_structure_field_eqn}
\end{flalign}
where in the last line we have substituted Eqs.~\eqref{eq:SI_probability_flux_ft}, \eqref{eq:SI_probability_flux_ft_avg}, and \eqref{eq:SI_conditional_probability}. Taylor-expanding $\hat{G}$ about $\bm{k} = \bm{0}$,
\begin{equation}
	\hat{G}(\bm{k}, \bm{r}, t)
	=
	g(\bm{r}, t)
	+
	\mathrm{i} \bm{k} \cdot \bm{d} (\bm{r}, t)
	+
	\cdots
	,
	\label{eq:SI_taylor_series}
\end{equation}
substituting the expansion into Eq.~\eqref{eq:SI_structure_field_eqn}, and collecting terms of like order in $\mathrm{i} \bm{k}$ yields the ordered set of equations,
\begin{flalign}
	\frac{\partial g(\bm{r}, t)}{\partial t}
	+
	\bm{u} (t) \cdot \bm{\nabla}_{\bm{r}} g
	-
	\frac{kT}{\gamma}
	\nabla_{\bm{r}}^{2} g
	-
	\frac{1}{\gamma}
	\bm{\nabla}_{\bm{r}} \cdot 
	[
	g \bm{\nabla}_{\bm{r}} V (\bm{r})
	]
	=
	0
	,
\end{flalign}
\begin{flalign}
	\frac{\partial \bm{d}(\bm{r}, t)}{\partial t}
	+ 
	\bm{u} (t) \cdot \bm{\nabla}_{\bm{r}} \bm{d}
	-
	\frac{kT}{\gamma}
	\nabla_{\bm{r}}^{2} \bm{d}
	-
	\frac{1}{\gamma}
	\bm{\nabla}_{\bm{r}} \cdot 
	[
	\bm{d}\bm{\nabla}_{\bm{r}} V (\bm{r})
	]^{\dag}
	=
	\frac{2kT}{\gamma} \bm{\nabla}_{\bm{r}} g
	+
	\frac{1}{\gamma}
	[
	g\bm{\nabla}_{\bm{r}} V (\bm{r})
	- 
	\braket{g \bm{\nabla}_{\bm{r}}V} g
	]
	.
\end{flalign}
The last two equations are exactly Eqs.~\eqref{eq:g_eqn} and \eqref{eq:d_eqn} from the main text. Conservation of probability requires the $g$- and $\bm{d}$-fields to satisfy periodic boundary conditions as well as the normalization conditions $\braket{g} = 1$ and $\braket{\bm{d}} = \bm{0}$.

\subsection{Derivation of Eqs. (6)-(7): effective drift velocity and diffusivity}

The effective drift velocity $\bm{U}(t)$ and diffusivity $\tens{D} (t)$ of the Brownian particle are related to the Fourier-transformed, average flux $\braket{\hat{\bm{J}}}$ via the large-wavelength expansion,
\begin{equation}
	\braket{\hat{\bm{J}}} (\bm{k}, t)
	=
	\hat{\rho}
	\left[  
		\bm{U} (t)
		-
		\mathrm{i} \bm{k} \cdot \tens{D} (t)
		+
		\cdots
	\right]
	.
	\label{eq:SI_average_flux_ft_eqn1}
\end{equation}
In order to derive expressions for $\bm{U}$ and $\tens{D}$, we insert Eqs.~\eqref{eq:SI_conditional_probability} and \eqref{eq:SI_taylor_series} into \eqref{eq:SI_probability_flux_ft_avg}, obtaining
\begin{flalign}
	\braket{\hat{\bm{J}}} (\bm{k}, t)
	&=
	\hat{\rho}
	\left(
	\bm{u} (t)
	-
	\frac{1}{\gamma}
	[
	kT \mathrm{i} \bm{k} 
	+
	\braket{\hat{G} \bm{\nabla}_{\bm{r}}V}
	]
	\right)
	\nonumber\\
	&=
	\hat{\rho}
	\left[
	\bm{u} (t) - \frac{1}{\gamma} \braket{g \bm{\nabla}_{\bm{r}}V}
	-
	\mathrm{i} \bm{k} 
	\left(
	\frac{kT}{\gamma} \tens{I} 
	+
	\frac{1}{\gamma}\braket{\bm{d} \bm{\nabla}_{\bm{r}}V}
	\right)
	+
	\cdots
	\right]
	.
	\label{eq:SI_average_flux_ft_eqn2}
\end{flalign}
Equating terms of like order in $\mathrm{i} \bm{k}$ in Eqs.~\eqref{eq:SI_average_flux_ft_eqn1} and \eqref{eq:SI_average_flux_ft_eqn2} furnishes the expressions,
\begin{equation}
	\bm{U}(t)
	=
	\bm{u} (t)
	-
	\frac{1}{\gamma} \braket{g \bm{\nabla}_{\bm{r}} V} (t)
	,
\end{equation}
\begin{equation}
	\tens{D}(t)
	=
	\frac{kT}{\gamma} \tens{I} 
	+
	\frac{1}{\gamma}\braket{\bm{d} \bm{\nabla}_{\bm{r}}V} (t)
	,
\end{equation}
which are exactly Eqs.~\eqref{eq:effective_drift}-\eqref{eq:effective_diffusivity} in the main text.

\section{3. Numerical Method} \label{sec:numerical_method}

\begin{figure}[!h]
	\begin{center}
		\includegraphics[width=0.7\linewidth]{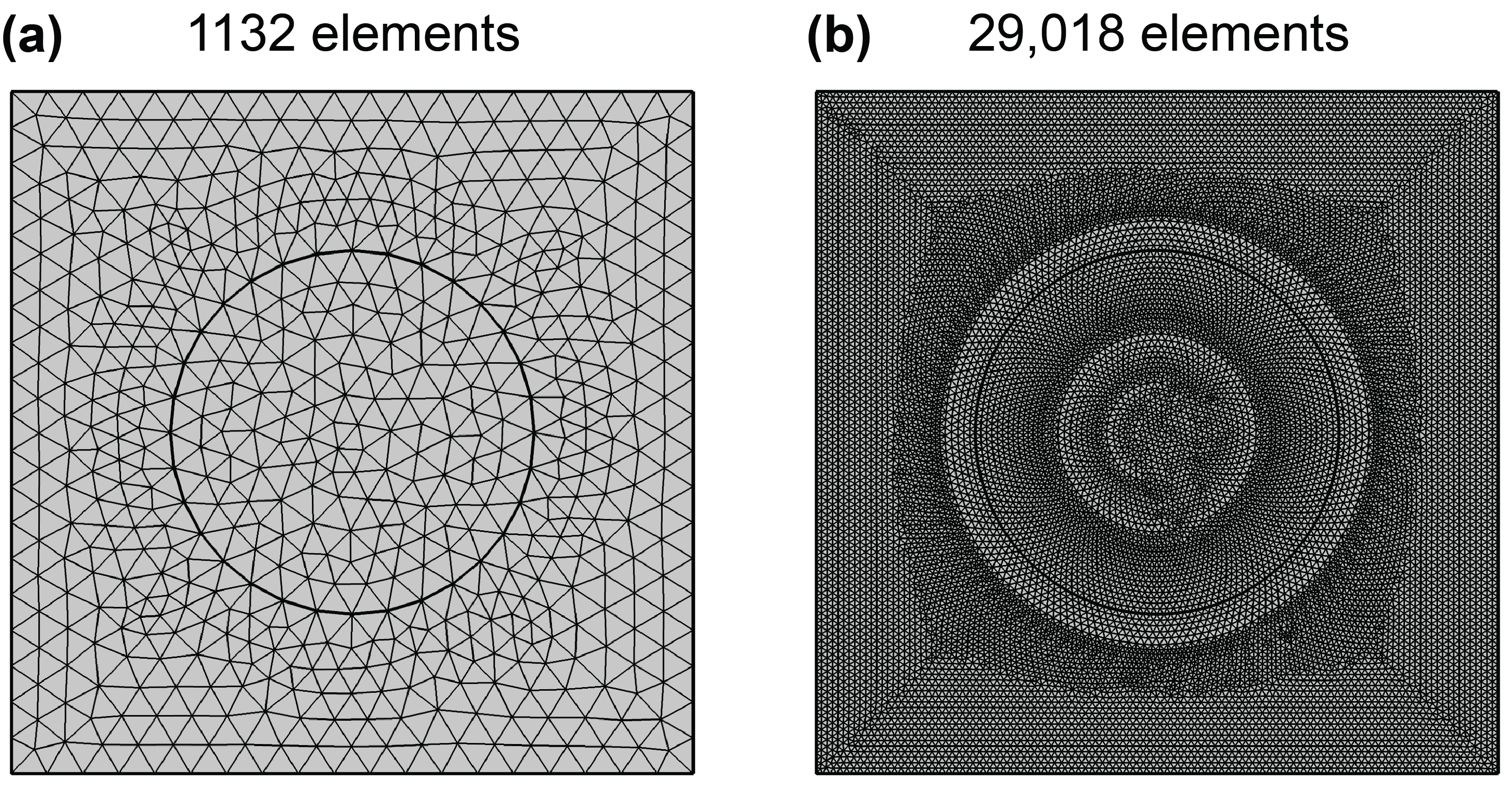}
		\caption{%
			Triangular meshes used for the finite-element calculations. Meshes containing (a) 1132 elements (for the time-dependent studies) and (b) 29,018 elements (for the stationary studies) were used. Coarser meshes were used in the time-dependent calculations to save computational time.
		}
		\label{Fig:SIMeshes}
	\end{center}
	\vspace{-18pt}
\end{figure}

Eqs.~\eqref{eq:g_eqn} and \eqref{eq:d_eqn} were solved using the finite-element method in COMSOL Multiphysics$^\text{\textregistered}$ (Version 5.5) with the ``Coefficient Form PDE'' physics interface. An $L \times L$ square cell was set up and discretized into triangular elements (Fig.~\ref{Fig:SIMeshes}). Periodic boundary conditions were applied to the $g$- and $\bm{d}$-fields at the edges of the cell. Studies were run using both time-dependent ($\bm{u}\not=\bm{0}$) and stationary ($\bm{u}=\bm{0}$) solvers. For the time-dependent studies, the $g$- and $\bm{d}$-fields were initialized to uniform values $1$ and $\bm{0}$, respectively, and time-advanced using the backward differentiation formula with a timestep $\Delta t = 0.001(2\uppi/\omega)$ until a periodic steady state was achieved. The number of periods needed to reach steady state generally increased with the oscillation frequency. For the stationary studies, the equations were solved iteratively using Newton's method and the normalization conditions $\braket{g} = 1$ and $\braket{\bm{d}} = \bm{0}$ were implemented as weak-form constraints. Upon solving for the $g$- and $\bm{d}$-fields, Eqs.~\eqref{eq:effective_drift} and \eqref{eq:effective_diffusivity} were evaluated using a fourth-order domain integration method and (in the time-dependent studies) subsequently time-averaged over the final oscillation period.

\section{4. Asymptotic Limits} \label{sec:asymptotic_limits}

\subsection{Derivation of Eq. (8): stationary traps with shallow potential wells}

If the harmonic traps held in a fixed configuration, $\bm{u} = \bm{0}$ and the $g$- and $\bm{d}$-fields achieve a steady state. Equations \eqref{eq:g_eqn} and \eqref{eq:d_eqn} then simplify to
\begin{flalign}
	kT
	\nabla_{\bm{r}}^{2} g (\bm{r})
	+
	\bm{\nabla}_{\bm{r}} \cdot 
	[
	g(\bm{r}) \bm{\nabla}_{\bm{r}} V (\bm{r})
	]
	=
	0
	,
	\label{eq:SI_g_eqn_steady}
\end{flalign}
\begin{flalign}
	kT
	\nabla_{\bm{r}}^{2} \bm{d}(\bm{r})
	+
	\bm{\nabla}_{\bm{r}} \cdot 
	[
	\bm{d}(\bm{r})\bm{\nabla}_{\bm{r}} V (\bm{r})
	]^{\dag}
	=
	-
	2kT \bm{\nabla}_{\bm{r}} g
	-
	g\bm{\nabla}_{\bm{r}} V 
	+ 
	\braket{g \bm{\nabla}_{\bm{r}}V} g
	.
	\label{eq:SI_d_eqn_steady}
\end{flalign}
Eq.~\eqref{eq:SI_g_eqn_steady} may be solved subject to the constraint $\braket{g} = 1$ to get the Boltzmann distribution,
\begin{equation}
	g (\bm{r})
	=
	\frac{\mathrm{e}^{-V(\bm{r})/kT}}{\braket{\mathrm{e}^{-V/kT}}}
	.
	\label{eq:SI_g_field_steady}
\end{equation}
The governing equation for the $\bm{d}$-field, Eq.~\eqref{eq:SI_d_eqn_steady}, then simplifies to
\begin{flalign}
	kT
	\nabla_{\bm{r}}^{2} \bm{d}
	+
	\bm{\nabla}_{\bm{r}} \cdot 
	(
	\bm{d}\bm{\nabla}_{\bm{r}} V 
	)^{\dag}
	=
	-
	kT \bm{\nabla}_{\bm{r}} g
	.
	\label{eq:SI_d_eqn_steady_2}
\end{flalign}

Eq.~\eqref{eq:SI_d_eqn_steady_2} cannot be solved analytically in general. However, for ``shallow'' potential wells, $\Delta V \ll kT$, we may Taylor-expand Eq.~\eqref{eq:SI_g_field_steady} as
\begin{equation}
	g
	=
	1 
	- 
	\frac{V - \braket{V}}{kT} 
	+
	\frac{V^{2} - \braket{V^{2}} - 2 \braket{V} (V - \braket{V})}{2(kT)^{2}} 
	+ 
	\cdots
	,
\end{equation}
so that Eq.~\eqref{eq:SI_d_eqn_steady_2} becomes
\begin{flalign}
	kT
	\nabla_{\bm{r}}^{2} \bm{d}
	+
	\bm{\nabla}_{\bm{r}} \cdot 
	(
	\bm{d}\bm{\nabla}_{\bm{r}} V 
	)^{\dag}
	=
	\left( 1 - \frac{V - \braket{V}}{kT} + \cdots \right)\bm{\nabla}_{\bm{r}} V
	.
	\label{eq:SI_d_eqn_steady_3}
\end{flalign}
To solve Eq.~\eqref{eq:SI_d_eqn_steady_3}, we expand the $\bm{d}$-field in a perturbation series,
\begin{equation}
	\bm{d} (\bm{r})
	=
	\bm{d}^{(0)} (\bm{r})
	+
	\bm{d}^{(1)} (\bm{r})
	+
	\cdots
	,
	\label{eq:SI_d_field_perturbation_series}
\end{equation}
where $\bm{d}^{(0)} = O(\Delta V/kT)$, $\bm{d}^{(1)} = O[\Delta V^{2}/(kT)^{2}]$, and so on. Inserting Eq.~\eqref{eq:SI_d_field_perturbation_series} into \eqref{eq:SI_d_eqn_steady_3} and collecting terms of like order in $\Delta V / kT$ yields the ordered set of equations,
\begin{flalign}
	kT
	\nabla_{\bm{r}}^{2} \bm{d}^{(0)}
	=
	\bm{\nabla}_{\bm{r}} V
	,
	\label{eq:SI_d_eqn_steady_4a}
\end{flalign}
\begin{flalign}
	kT
	\nabla_{\bm{r}}^{2} \bm{d}^{(1)}
	=
	-
	\bm{\nabla}_{\bm{r}} \cdot 
	(
	\bm{d}^{(0)}\bm{\nabla}_{\bm{r}} V 
	)^{\dag}
	-
	\frac{V - \braket{V}}{kT} \bm{\nabla}_{\bm{r}} V
	,
	\label{eq:SI_d_eqn_steady_4b}
\end{flalign}
subject to the constraints $\braket{\bm{d}^{(0)}} = \bm{0}$, $\braket{\bm{d}^{(1)}} = \bm{0}$, etc. Since $V$ and $\bm{d}$ are spatially periodic, Eqs.~\eqref{eq:SI_d_eqn_steady_4a}-\eqref{eq:SI_d_eqn_steady_4b} may be sequentially solved by means of Fourier series:
\begin{equation}
	\bm{d}^{(0)} (\bm{r})
	=
	-
	\frac{1}{kT}
	\sum_{\bm{q} \not= \bm{0}}
	\frac{\mathrm{i} \bm{q}}{q^{2}} 
	V_{\bm{q}} \mathrm{e}^{\mathrm{i} \bm{q} \cdot \bm{r}}
	,
	\label{eq:SI_d0_soln}
\end{equation}
\begin{equation}
	\bm{d}^{(1)} (\bm{r})
	=
	\frac{1}{2(kT)^{2}}
	\sum_{\bm{q} \not= \bm{0}}
	\sum_{\bm{q}' \not= \bm{0}}
	\cdot
	\left( \frac{\mathrm{i} \bm{q}}{q^{2}} + \frac{2 \mathrm{i} \bm{q} \cdot (\bm{q} - \bm{q}')\bm{q}'}{q^{2} q'^{2}} \right)
	V_{\bm{q} - \bm{q}'} V_{\bm{q}'}
	\mathrm{e}^{\mathrm{i} \bm{q} \cdot \bm{r}}
	,
	\label{eq:SI_d1_soln}
\end{equation}
where $\bm{q}$ is the discrete wavevector and $V_{\bm{q}} \equiv L^{-2} \int_{L^{2}} [V(\bm{r}) - \braket{V}] \mathrm{e}^{- \mathrm{i} \bm{q} \cdot \bm{r}} \, \mathrm{d} \bm{r}$ denotes the Fourier integral of $V$. 

By use of Eqs.~\eqref{eq:effective_diffusivity} and \eqref{eq:SI_d_field_perturbation_series}, the effective diffusivity of the Brownian particle is given by
\begin{flalign}
	\tens{D}
	&=
	\frac{kT}{\gamma} \tens{I}
	+
	\frac{1}{\gamma} \braket{\bm{d} \bm{\nabla}_{\bm{r}} V}
	\nonumber\\
	&=
	\frac{kT}{\gamma} \tens{I}
	+
	\frac{1}{\gamma} \braket{\bm{d}^{(0)} \bm{\nabla}_{\bm{r}} V}
	+
	\frac{1}{\gamma} \braket{\bm{d}^{(1)} \bm{\nabla}_{\bm{r}} V}
	+
	\cdots
	.
	\label{eq:SI_diffusivity_tensor_smallV_expansion}
\end{flalign}
Multiplying Eqs.~\eqref{eq:SI_d0_soln} by $\bm{\nabla}_{\bm{r}} V = \sum_{\bm{q}\not=\bm{0}} \mathrm{i} \bm{q} V_{\bm{q}} \mathrm{e}^{\mathrm{i} \bm{q} \cdot \bm{r}}$ and averaging over an $L \times L$ cell yields the force-displacement dyads,
\begin{equation}
	\braket{\bm{d}^{(0)} \bm{\nabla}_{\bm{r}} V}
	=
	-
	\frac{1}{kT}
	\sum_{\bm{q} \not= \bm{0}}
	\frac{\bm{q}\bm{q}}{q^{2}} 
	|V_{\bm{q}}|^{2}
	,
	\label{eq:SI_d0_gradV_avg}
\end{equation}
\begin{equation}
	\braket{\bm{d}^{(1)} \bm{\nabla}_{\bm{r}} V}
	=
	\frac{1}{2(kT)^{2}}
	\sum_{\bm{q} \not= \bm{0}}
	\sum_{\bm{q}' \not= \bm{0}}
	\left( \frac{\bm{q}\bm{q}}{q^{2}} + \frac{2 \bm{q} \cdot (\bm{q} - \bm{q}')\bm{q}'\bm{q}}{q^{2} q'^{2}} \right)
	V_{\bm{q} - \bm{q}'} V_{\bm{q}'} V_{-\bm{q}}
	,
	\label{eq:SI_d1_gradV_avg}
\end{equation}
where $|V_{\bm{q}}|^{2} \equiv V_{\bm{q}} V_{-\bm{q}}$. Thus, the diffusivity tensor $\tens{D}$ admits the Fourier-series representation,
\begin{flalign}
	\tens{D}
	&=
	\frac{kT}{\gamma} \,
	\left[
	\tens{I}
	-
	\frac{1}{(kT)^{2}}\sum_{\bm{q} \not= \bm{0}}
	\frac{\bm{q}\bm{q}}{q^{2}} 
	|V_{\bm{q}}|^{2}
	+
	\frac{1}{2(kT)^{3}}
	\sum_{\bm{q} \not= \bm{0}}
	\sum_{\bm{q}' \not= \bm{0}}
	\left( \frac{\bm{q}\bm{q}}{q^{2}} + \frac{2 \bm{q} \cdot (\bm{q} - \bm{q}')\bm{q}'\bm{q}}{q^{2} q'^{2}} \right)
	V_{\bm{q} - \bm{q}'} V_{\bm{q}'} V_{-\bm{q}}
	+
	\cdots
	\right]
	.
	\label{eq:SI_diffusivity_smallV_1}
\end{flalign}

An alternative expression for $\tens{D}$ can be obtained by writing leading-order displacement field as the negative gradient of a potential, 
\begin{equation}
	\bm{d}^{(0)} (\bm{r})
	=
	- \frac{1}{kT} \bm{\nabla}_{\bm{r}} \varPhi(\bm{r})
	,
	\label{eq:SI_diffusivity_as_gradient_of_potential}
\end{equation}
where $\varPhi(\bm{r})$ satisfies the Poisson equation,
\begin{equation}
	\nabla^{2}_{\bm{r}} \varPhi(\bm{r}) = - [V(\bm{r}) - \braket{V}],
	\label{eq:SI_poisson_eqn}
\end{equation}
subject to the closure $\braket{\varPhi} = 0$. The Fourier-series solution of Eq.~\eqref{eq:SI_poisson_eqn} is
\begin{equation}
	\varPhi(\bm{r}) 
	=
	\sum_{\bm{q} \not= \bm{0}}
	q^{-2}
	V_{\bm{q}} \mathrm{e}^{\mathrm{i} \bm{q} \cdot \bm{r}}
	.
	\label{eq:SI_poisson_field_soln}
\end{equation}
By use of Eqs.~\eqref{eq:SI_d0_soln}, \eqref{eq:SI_d1_gradV_avg}, and the convolution theorem, it can be shown that
\begin{equation}
	\braket{\bm{d}^{(1)} \bm{\nabla}_{\bm{r}} V}
	=
	\frac{1}{2kT} 
	\braket{(V - \braket{V})^{2} \bm{\nabla}_{\bm{r}} \bm{d}^{(0)}}
	+
	\braket{(\bm{\nabla}_{\bm{r}} \bm{d}^{(0)} \cdot \bm{\nabla}_{\bm{r}} V)\bm{d}^{(0)}} 
	.
	\label{eq:SI_d1_gradV_avg_identity}
\end{equation}
Then, by Eqs.~\eqref{eq:SI_d0_gradV_avg}, \eqref{eq:SI_diffusivity_as_gradient_of_potential}, \eqref{eq:SI_poisson_field_soln}, and \eqref{eq:SI_d1_gradV_avg_identity}, it follows that
\begin{equation}
	\braket{\bm{d}^{(0)} \bm{\nabla}_{\bm{r}} V}
	=
	- \frac{1}{kT} \braket{\bm{\nabla}_{\bm{r}} \varPhi \bm{\nabla}_{\bm{r}} V}
	,
	\label{eq:SI_d0_gradV_avg_2}
\end{equation}
\begin{equation}
	\braket{\bm{d}^{(1)} \bm{\nabla}_{\bm{r}} V}
	=
	\frac{1}{2(kT)^{2}} 
	\left[
	-
	\braket{(V - \braket{V})^{2} \bm{\nabla}_{\bm{r}} \bm{\nabla}_{\bm{r}} \varPhi}
	+
	2 \braket{(\bm{\nabla}_{\bm{r}} \bm{\nabla}_{\bm{r}} \varPhi \cdot \bm{\nabla}_{\bm{r}} V)\bm{\nabla}_{\bm{r}} \varPhi} 
	\right]
	.
	\label{eq:SI_d1_gradV_avg_2}
\end{equation}
Substituting Eqs.~\eqref{eq:SI_d0_gradV_avg_2}-\eqref{eq:SI_d1_gradV_avg_2} into \eqref{eq:SI_diffusivity_tensor_smallV_expansion} then gives the alternative representation,
\begin{flalign}
	\tens{D}
	&=
	\frac{kT}{\gamma} 
	\left(
	\tens{I}
	-
	\frac{1}{(kT)^{2}}
	\braket{\bm{\nabla}_{\bm{r}} \varPhi \bm{\nabla}_{\bm{r}} V}
	+
	\frac{1}{2(kT)^{3}}
	\left[
	-
	\braket{(V - \braket{V})^{2} \bm{\nabla}_{\bm{r}} \bm{\nabla}_{\bm{r}} \varPhi}
	+
	2 \braket{(\bm{\nabla}_{\bm{r}} \bm{\nabla}_{\bm{r}} \varPhi \cdot \bm{\nabla}_{\bm{r}} V)\bm{\nabla}_{\bm{r}} \varPhi} 
	\right]
	+
	\cdots
	\right)
	.
	\label{eq:SI_diffusivity_smallV_2}
\end{flalign}

Since $V(\bm{r})$ is isotropic, only the trace of the steady diffusivity tensor need be computed: $D \equiv \tfrac{1}{2} \tens{D} : \tens{I}$. Using Eq.~\eqref{eq:SI_diffusivity_tensor_smallV_expansion}, the scalar diffusivity $D$ is given by
\begin{equation}
	D
	=
	\frac{kT}{\gamma}
	+
	\frac{1}{2\gamma} \braket{\bm{d}^{(0)} \cdot \bm{\nabla}_{\bm{r}} V}
	+
	\frac{1}{2\gamma} \braket{\bm{d}^{(1)} \cdot \bm{\nabla}_{\bm{r}} V}
	+
	\cdots
	.
	\label{eq:SI_diffusivity_scalar_smallV_expansion}
\end{equation}
Taking the trace of Eqs.~\eqref{eq:SI_d0_gradV_avg_2}-\eqref{eq:SI_d1_gradV_avg_2}, integrating by parts, and applying Eq.~\eqref{eq:SI_poisson_eqn} then gives
\begin{flalign}
	\braket{\bm{d}^{(0)} \cdot \bm{\nabla}_{\bm{r}} V}
	&=
	- \frac{1}{kT} \braket{\bm{\nabla}_{\bm{r}} \varPhi \cdot \bm{\nabla}_{\bm{r}} V}
	\nonumber\\
	&=
	-\frac{1}{kT} \braket{(V - \braket{V})^{2}}
	,
	\label{eq:SI_d0_gradV_avg_3}
\end{flalign}
\begin{flalign}
	\braket{\bm{d}^{(1)} \cdot \bm{\nabla}_{\bm{r}} V}
	&=
	\frac{1}{2(kT)^{2}} 
	\left[
	-
	\braket{(V - \braket{V})^{2} \nabla^{2}_{\bm{r}} \varPhi}
	+
	2 \braket{(\bm{\nabla}_{\bm{r}} \bm{\nabla}_{\bm{r}} \varPhi \cdot \bm{\nabla}_{\bm{r}} V) \cdot\bm{\nabla}_{\bm{r}} \varPhi} 
	\right]
	\nonumber\\
	&=
	\frac{1}{2(kT)^{2}} 
	\left[
	\braket{(V - \braket{V})^{3}}
	+
	\braket{\bm{\nabla}_{\bm{r}} (|\bm{\nabla}_{\bm{r}} \varPhi|^{2}) \cdot \bm{\nabla}_{\bm{r}} V} 
	\right]
	.
	\label{eq:SI_d1_gradV_avg_3}
\end{flalign}
Inserting Eqs.~\eqref{eq:SI_d0_gradV_avg_3}-\eqref{eq:SI_d1_gradV_avg_3} into \eqref{eq:SI_diffusivity_scalar_smallV_expansion} then gives
\begin{equation}
	D
	=
	\frac{kT}{\gamma}
	\left(
	1
	-
	\frac{1}{2(kT)^{2}}
	\braket{(V - \braket{V})^{2}}
	+
	\frac{1}{4(kT)^{3}}
	\left[
	\braket{(V - \braket{V})^{3}}
	+
	\braket{\bm{\nabla}_{\bm{r}} (|\bm{\nabla}_{\bm{r}} \varPhi|^{2}) \cdot \bm{\nabla}_{\bm{r}} V} 
	\right]
	+
	\cdots
	\right)
	.
	\label{eq:SI_diffusivity_scalar_smallV_expansion}
\end{equation}
The last expression is exactly Eq.~\eqref{eq:diffusivity_soft_traps} from the main text.

\subsection{Derivation of Eq. (9): stationary traps with deep potential wells}

\begin{figure}[!h]
	\begin{center}
		\includegraphics[width=1\linewidth]{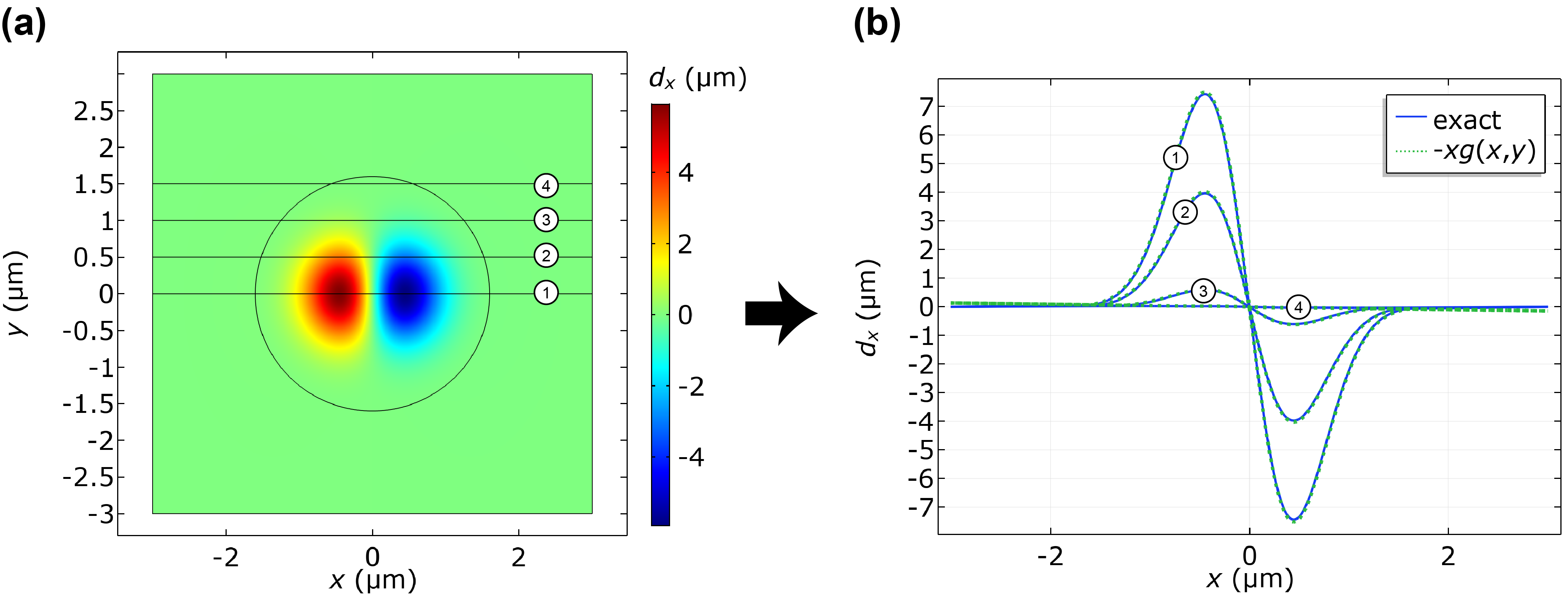}
		\caption{
		Comparison of numerical solution for the steady displacement field density $d_{x}(x,y)$ against the particular solution [see Eq.~\eqref{eq:SI_d_field_largeV_approx}] for a stiff trap, $\kappa=5$ $kT/\mu$m$^2$. (a) 2D contour plot of $d_{x}$ with line traces at four distinct values of $y$. (b) Plot of $d_{x}$ against $x$ for each line trace shows favorable agreement to Eq.~\eqref{eq:SI_d_field_largeV_approx}. 
		}
		\label{Fig:SI_StrongTrap_dx_vs_x}
	\end{center}
	\vspace{-18pt}
\end{figure}

For stationary, ``deep'' potential wells, $\Delta V \gg kT$, the small-potential perturbation series \eqref{eq:SI_d_field_perturbation_series} fails to converge. Unfortunately, no exact analytical solution of Eq.~\eqref{eq:SI_d_eqn_steady_2} is readily available. However, one can take advantage of the fact that, for deep potential wells, the probability density is strongly localized near the origin $\bm{r} = \bm{0}$ of the lattice cell where the potential-energy field $V(\bm{r})$ is minimized. Then, a useful {\it approximation} of the $\bm{d}$-field is
\begin{flalign}
	\bm{d} (\bm{r})
	&\approx
	- \bm{r} g (\bm{r})
	\nonumber\\
	&=
	- \frac{\bm{r} \mathrm{e}^{-V(\bm{r})/kT}}{\braket{\mathrm{e}^{-V/kT}}}
	.
	\label{eq:SI_d_field_largeV_approx}
\end{flalign}
Eq.~\eqref{eq:SI_d_field_largeV_approx} is the particular solution of Eq.~\eqref{eq:SI_d_eqn_steady_2} and conserves probability, $\braket{\bm{d}} = \bm{0}$. However, this particular solution clearly violates the periodic boundary conditions at the edges of the lattice cell $x = \pm L/2$, $y=\pm L/2$, incurring an error of $O(L\mathrm{e}^{-\Delta V/kT}/\braket{\mathrm{e}^{-V/kT}})$ that decreases in magnitude with increasing trap stiffness. Fig.~\ref{Fig:SI_StrongTrap_dx_vs_x} compares the approximation, Eq.~\eqref{eq:SI_d_field_largeV_approx}, against the ``exact'' numerical solution for the displacement field, showing very good agreement. The slight error in the approximation is due to the neglect of the homogeneous solution of Eq.~\eqref{eq:SI_d_eqn_steady_2}, which is complicated by the 2D potential-energy field given by Eq.~\eqref{eq:harmonic_potential}. It will be shown that the error in this approximation for the $\bm{d}$-field quantitatively (though not qualitatively) impacts the prediction for the effective diffusivity.

\begin{figure}[!h]
	\begin{center}
		\includegraphics[width=0.5\linewidth]{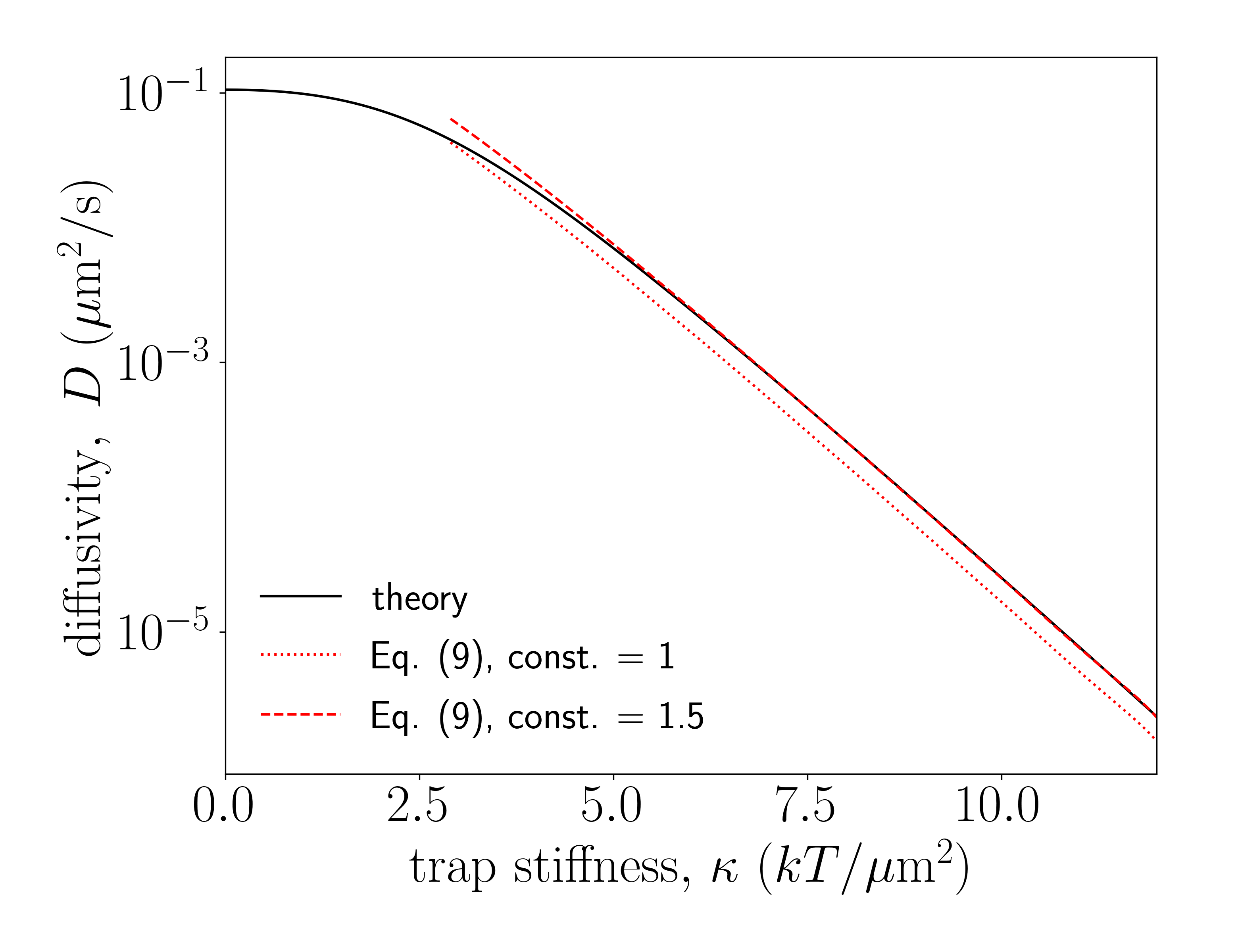}
		\caption{%
		Log-linear plot of diffusivity $D$ against trap stiffness $\kappa$. The full numerical solution (solid curve) is compared against Eq.~\eqref{eq:diffusivity_stiff_traps} (dashed curves) using two different constants of proportionality. Irrespective of the numerical prefactor, Eq.~\eqref{eq:diffusivity_stiff_traps} demonstrates the appropriate scaling with the trapping strength and is consistent with Kramers' theory of activated escape. A proportionality constant of 1.5 gives quantitative agreement with the exact solution for the specific geometry considered in this study.
		}
		\label{Fig:SI_StrongTrap_D_vs_K_LogLinearPlot}
	\end{center}
	\vspace{-18pt}
\end{figure}

Using Eq.~\eqref{eq:harmonic_potential} for $V(\bm{r})$ and Eq.~\eqref{eq:SI_d_field_largeV_approx} for $\bm{d}(\bm{r})$, the force-displacement dyad that appears in Eq.~\eqref{eq:diffusivity_soft_traps} can now be approximated as
\begin{flalign}
	\braket{\bm{d} \bm{\nabla}_{\bm{r}} V}
	&\approx
	-
	\frac{1}{\kappa}
	\frac{\braket{\mathrm{e}^{-V/kT}\bm{\nabla}_{\bm{r}} V \bm{\nabla}_{\bm{r}} V}}{\braket{\mathrm{e}^{-V/kT}}}
	,
	\label{eq:SI_d_gradV_avg_largeV}
\end{flalign}
where we've used the fact that $\bm{\nabla}_{\bm{r}} V = \kappa \bm{r}$ for $r \le \frac{1}{2} W_{\text{trap}}$ and $=\bm{0}$ otherwise. Defining the well depth as $\Delta V =\tfrac{1}{8} \kappa W_{\text{trap}}^{2}$, the cell averages in Eq.~\eqref{eq:SI_d_gradV_avg_largeV} become
\begin{flalign}
	\braket{\mathrm{e}^{-V/kT}}
	&=
	\frac{2\uppi kT}{\kappa L^{2}}\left( 1 - \mathrm{e}^{-\Delta V/kT} \right)
	+
	\bigg( 1- \frac{2\uppi \Delta V}{\kappa L^{2}} \bigg) \mathrm{e}^{-\Delta V/kT}
	,
\end{flalign}
\begin{flalign}
	\braket{\mathrm{e}^{-V/kT}\bm{\nabla}_{\bm{r}} V \bm{\nabla}_{\bm{r}} V}
	&=
	\frac{2\uppi (kT)^{2}}{L^{2}}
	\left[
		1
		-
		\left( 1 + \frac{\Delta V}{kT} \right)\mathrm{e}^{-\Delta V/kT}
	\right]
	\tens{I}
	.
\end{flalign}
Substitution into Eq.~\eqref{eq:SI_d_gradV_avg_largeV} then gives, upon simplification,
\begin{flalign}
	\braket{\bm{d} \bm{\nabla}_{\bm{r}} V}
	&\approx
	kT
	\left\{
		-1
		+
		\left[  
		1
		+
		\frac{2\uppi kT}{\kappa L^{2}}
		\left(
			\mathrm{e}^{\Delta V/kT}
			-
			\frac{\Delta V}{kT}
			-
			1
		\right)
		\right]^{-1}
	\right\}
	\tens{I}
	\nonumber\\
	&\approx
	\left(
	- k T
	+
	\frac{\kappa L^{2}}{2\uppi}
	\mathrm{e}^{-\Delta V / kT} 
	\right)
	\tens{I}
	\quad
	\text{for}
	\quad
	\Delta V \gg kT.
\end{flalign}
Substitution into Eq.~\eqref{eq:effective_diffusivity} and replacing $\kappa \tens{I}$ by $(\bm{\nabla}_{\bm{r}}\bm{\nabla}_{\bm{r}}V)|_{\bm{r}=\bm{0}}$ then gives the following approximation for the diffusivity tensor:
\begin{equation}
	\tens{D}
	\approx
	\frac{L^{2}}{2\uppi \gamma}
	\mathrm{e}^{-\Delta V / kT}
	(\bm{\nabla}_{\bm{r}}\bm{\nabla}_{\bm{r}}V)|_{\bm{r}=\bm{0}}
	,
\end{equation}
or, upon taking one-half the trace,
\begin{equation}
	D
	\approx
	\frac{L^{2}}{4\uppi \gamma}
	\mathrm{e}^{-\Delta V / kT}
	(\nabla_{\bm{r}}^{2}V)|_{\bm{r}=\bm{0}}
	.
	\label{eq:SI_diffusivity_largeV_approx}
\end{equation}
This is exactly the form that would be predicted by Kramers' theory for the escape of a Brownian particle from a deep potential well \cite{kramers1940brownian,brinkman1956brownian,brinkman1956brownian2}.
Comparison of Eq.~\eqref{eq:SI_diffusivity_largeV_approx} to numerical calculations of $D$ indicates the qualitatively correct dependence on the trapping strength, but quantitative discrepancies due to errors in the approximation \eqref{eq:SI_d_field_largeV_approx} for the $\bm{d}$-field (see Fig.~\ref{Fig:SI_StrongTrap_D_vs_K_LogLinearPlot}). Quantitative agreement can be obtained by renormalizing the above result by a factor that depends upon the ratio $W_{\text{trap}} / L$. Therefore, we write
\begin{equation}
	D
	\propto
	\frac{L^{2}}{4\uppi \gamma}
	\mathrm{e}^{-\Delta V / kT}
	(\nabla_{\bm{r}}^{2}V)|_{\bm{r}=\bm{0}}
	\label{eq:SI_diffusivity_largeV_approx_proportionality}
\end{equation}
up to a proportionality constant. Eq.~\eqref{eq:SI_diffusivity_largeV_approx_proportionality} is identical to Eq.~\eqref{eq:diffusivity_stiff_traps} from the main text.
For traps of diameter $W_{\text{trap}} = 3.2$ $\mu$m spaced a distance $L=6$ $\mu$m apart, a proportionality constant of 1.5 gives quantitative agreement with the exact dispersion theory (see Fig.~\ref{Fig:SI_StrongTrap_D_vs_K_LogLinearPlot}).

\subsection{Derivation of Eq. (11): oscillating traps in the high-frequency limit}

In the high-frequency limit, the potential-energy field is cycled in the $x$-direction at a rate much faster than the response time of the Brownian particle. A reasonable model for this system is a quasi-steady, uniform convection in the $x$-direction, for which we make the ansatz $g = g(y)$ and $d_{y} = d_{y} (y)$ (for the time being, we will ignore the $d_{x}$-field). Eqs.~\eqref{eq:g_eqn} and \eqref{eq:d_eqn} then simplify to
\begin{equation}
	kT
	\frac{\mathrm{d}^{2}g}{\mathrm{d} y^{2}}
	+
	\frac{\partial V}{\partial y} \frac{\mathrm{d} g}{\mathrm{d} y}
	+
	\left( \frac{\partial^{2} V}{\partial x^{2}} + \frac{\partial^{2} V}{\partial y^{2}} \right) g
	=
	0
	,
	\label{eq:SI_g_eqn_1d}
\end{equation}
\begin{equation}
	kT
	\frac{\mathrm{d}^{2} d_{y}}{\mathrm{d} y^{2}}
	+
	\frac{\partial V}{\partial y} \frac{\mathrm{d} d_{y}}{\mathrm{d} y}
	+
	\left( \frac{\partial^{2} V}{\partial x^{2}} + \frac{\partial^{2} V}{\partial y^{2}} \right) d_{y}
	=
	-
	2 kT \frac{\mathrm{d} g}{\mathrm{d} y}
	- 
	g \frac{\partial V}{\partial y}
	+
	\bigg\langle g \frac{\partial V}{\partial y} \bigg\rangle g
	.
	\label{eq:SI_dy_eqn_1d}
\end{equation}
Averaging Eqs.~\eqref{eq:SI_g_eqn_1d}-\eqref{eq:SI_dy_eqn_1d} over the $x$-direction only and defining the modified potential,
\begin{equation}
	v(y) 
	=
	\frac{1}{L} \int_{-L/2}^{L/2} V(x,y) \, \mathrm{d} x
	,
\end{equation}
then gives
\begin{equation}
	kT
	\frac{\mathrm{d}^{2}g}{\mathrm{d} y^{2}}
	+
	\frac{\mathrm{d}}{\mathrm{d} y} \left( g\frac{\partial v}{\partial y} \right)
	=
	0
	,
	\label{eq:SI_g_eqn_1d_2}
\end{equation}
\begin{equation}
	kT
	\frac{\mathrm{d}^{2} d_{y}}{\mathrm{d} y^{2}}
	+
	\frac{\mathrm{d}}{\mathrm{d} y} \left( d_{y} \frac{\partial v}{\partial y} \right)
	=
	-
	2 kT \frac{\mathrm{d} g}{\mathrm{d} y}
	- 
	g \frac{\mathrm{d} v}{\mathrm{d} y}
	+
	\bigg\langle g \frac{\mathrm{d} v}{\mathrm{d} y} \bigg\rangle g
	,
	\label{eq:SI_dy_eqn_1d_2}
\end{equation}
where we have applied the conditions $V(L/2,y)=V(-L/2,y)$ and $(\partial V / \partial x)|_{x = \pm L/2} = 0$. Here, it is understood that the cell average of a one-dimensional (1D) function $f(y)$ simplifies to a 1D average in the $y$-direction, $\braket{f} = L^{-1} \int_{-L/2}^{L/2} f(y)\, \mathrm{d}y$.

Eqs.~\eqref{eq:SI_g_eqn_1d_2}-\eqref{eq:SI_dy_eqn_1d_2} are the 1D versions of Eqs.~\eqref{eq:SI_g_eqn_steady}-\eqref{eq:SI_d_eqn_steady}. The solution of Eq.~\eqref{eq:SI_g_eqn_1d_2} for the $g$-field, subject to the constraint $\braket{g} = 1$, is the 1D analog of Eq.~\eqref{eq:SI_g_field_steady}:
\begin{equation}
	g(y)
	=
	\frac{\mathrm{e}^{-v(y)/kT}}{\braket{\mathrm{e}^{-v/kT}}}
	.
\end{equation}
Eq.~\eqref{eq:SI_dy_eqn_1d_2} then simplifies to
\begin{flalign}
	kT
	\frac{\mathrm{d}^{2} d_{y}}{\mathrm{d} y^{2}}
	+
	\frac{\mathrm{d}}{\mathrm{d} y} \left( d_{y} \frac{\partial v}{\partial y} \right)
	&=
	-
	kT \frac{\mathrm{d} g}{\mathrm{d} y}
	\nonumber \\
	&=
	\frac{\mathrm{e}^{-v/kT}}{\braket{\mathrm{e}^{-v/kT}}}
	\frac{\mathrm{d} v}{\mathrm{d} y}
	,
	\label{eq:SI_dy_eqn_1d_3}
\end{flalign}
which is the 1D analog of Eq.~\eqref{eq:SI_d_eqn_steady_2}. Unlike the 2D problem, the 1D problem admits an exact analytical solution:
\begin{flalign}
	d_{y} (y)
	&=
	- y g (y)
	+
	c_{1} \mathrm{e}^{- v(y)/kT}
	\int_{0}^{y}
	\mathrm{e}^{v(\eta)/kT} \, \mathrm{d} \eta
	+
	c_{2} L \mathrm{e}^{-v(y) / kT}
	\nonumber\\
	&=
	- \frac{y \mathrm{e}^{-v(y)/kT}}{\braket{\mathrm{e}^{-v/kT}}} 
	+
	c_{1} \mathrm{e}^{- v(y)/kT}
	\int_{0}^{y}
	\mathrm{e}^{v(\eta)/kT} \, \mathrm{d} \eta
	+
	c_{2} L \mathrm{e}^{-v(y) / kT}
	.
	\label{eq:SI_dy_field_1d_general_soln}
\end{flalign}
The first term on the right-hand side of Eq.~\eqref{eq:SI_dy_field_1d_general_soln} is simply the particular solution of Eq.~\eqref{eq:SI_dy_eqn_1d_3}; it is the 1D analog of Eq.~\eqref{eq:SI_d_field_largeV_approx}, which was used to approximate the full solution in the strong-potential limit. The remaining terms in Eq.~\eqref{eq:SI_dy_field_1d_general_soln} are the homogeneous solutions, with constants $c_{1}$, $c_{2}$ that must be determined from the periodicity and normalization conditions,
\begin{subequations}
\begin{gather}
	d_{y} (L/2) - d_{y} (-L/2) = 0,
	\label{eq:SI_dy_1d_bc1}
	\\
	\braket{d_{y}}
	=
	\frac{1}{L} \int_{-L/2}^{L/2} d_{y}(y) \, \mathrm{d}y
	=
	0.
	\label{eq:SI_dy_1d_bc2}
\end{gather}
	\label{eq:SI_dy_1d_bcs}%
\end{subequations}
Inserting Eq.~\eqref{eq:SI_dy_field_1d_general_soln} into \eqref{eq:SI_dy_1d_bcs}, setting $v(L/2) = v(-L/2)$, and solving for the two unknowns $c_{1}$ and $c_{2}$ gives
\begin{subequations}
\begin{flalign}
	c_{1} &= \braket{\mathrm{e}^{-v/kT}}^{-1} \braket{\mathrm{e}^{v/kT}}^{-1}
	,
	\\
	c_{2} &= 
	\frac{1}{L} \braket{\mathrm{e}^{-v/kT}}^{-2}
	\left(
		\braket{y\mathrm{e}^{-v/kT}}
		-
		\braket{\mathrm{e}^{v/kT}}^{-1}
		\bigg\langle \mathrm{e}^{-v/kT}\int_{0}^{y} \mathrm{e}^{v(\eta)/kT} \mathrm{d} \eta \bigg\rangle
	\right)
	.
\end{flalign}
	\label{eq:SI_dy_1d_constants}%
\end{subequations}

With the solution for $d_{y}(y)$ fully specified, it remains to compute the effective diffusivity along the $y$-axis. Multiplying Eq.~\eqref{eq:SI_dy_field_1d_general_soln} by $\mathrm{d} v / \mathrm{d}y$, applying the inverse chain rule, and averaging over the $y$-direction gives
\begin{flalign}
	\bigg\langle d_{y} \frac{\mathrm{d} v}{\mathrm{d}y}\bigg\rangle
	&=
	kT
	\left(
	\braket{\mathrm{e}^{-v/kT}}^{-1}
	\bigg\langle y \frac{\mathrm{d}\mathrm{e}^{-v/kT}}{\mathrm{d}y} \bigg\rangle 
	-
	c_{1} 
	\bigg\langle \frac{\mathrm{d}\mathrm{e}^{-v/kT}}{\mathrm{d}y}
	\int_{0}^{y}
	\mathrm{e}^{v(\eta)/kT} \, \mathrm{d} \eta
	\bigg\rangle 
	-
	c_{2} L 
	\bigg\langle \frac{\mathrm{d}\mathrm{e}^{-v/kT}}{\mathrm{d}y} \bigg\rangle 
	\right).
	\label{eq:SI_dy_gradv_1d_avg}
\end{flalign}
Inserting Eqs.~\eqref{eq:SI_dy_1d_constants} into \eqref{eq:SI_dy_gradv_1d_avg} and integrating by parts then gives, after some simplification,
\begin{flalign}
	\bigg\langle d_{y} \frac{\mathrm{d} v}{\mathrm{d}y}\bigg\rangle
	&=
	k T
	\left( 
		-1
		+
		\braket{\mathrm{e}^{-v/kT}}^{-1} \braket{\mathrm{e}^{v/kT}}^{-1}
	\right)
	.
\end{flalign}
Since $d_{y}$ is independent of $x$, $\braket{d_{y} (\partial V / \partial y)} = \braket{d_{y} (\mathrm{d} v / \mathrm{d} y)}$. Thus, the $yy$ component of Eq.~\eqref{eq:effective_diffusivity} simplifies to
\begin{flalign}
	\overline{D}_{yy}
	&=
	\frac{kT}{\gamma}
	+
	\frac{1}{\gamma}
	\bigg\langle d_{y} \frac{\mathrm{d} v}{\mathrm{d}y}\bigg\rangle
	\nonumber\\
	&=
	\frac{kT}{\gamma}
	\braket{\mathrm{e}^{-v/kT}}^{-1} \braket{\mathrm{e}^{v/kT}}^{-1}
	,
	\label{eq:SI_Dyy_1d}
\end{flalign}
where an overbar is used to denote the long-time average over one periodic cycle.
This is the classical result for diffusion of a Brownian particle in a 1D periodic potential \cite{lifson1962self,Festa1978}.

Up until now, we have neglected the $d_{x}$-field, which appears in the $xx$-component of Eq.~\eqref{eq:effective_diffusivity} and, therefore, influences the effective diffusivity along the $x$-axis. To a first approximation, we assume that the gradients in the $x$-direction have been ``smeared out'' so that dispersion in that direction is negligible: $\braket{d_{x} (\partial V / \partial x)} \approx 0$. This approximation is consistent with a model of dispersion in an effectively 1D potential. Therefore, the $xx$-component of Eq.~\eqref{eq:effective_diffusivity} (time-averaged) is simply the Stokes-Einstein-Sutherland diffusivity:
\begin{equation}
	\overline{D}_{xx} = \frac{kT}{\gamma}
	.
	\label{eq:SI_Dxx_1d}
\end{equation}
Eqs.~\eqref{eq:SI_Dyy_1d} and \eqref{eq:SI_Dxx_1d} are exactly the same as Eq.~\eqref{eq:diffusivity_high_frequency} from the main text.

\section{5. Brownian Dynamics Simulations} \label{sec:brownian_dynamics_simulations}

The Langevin equation of motion corresponding to Eqs.~\eqref{eq:SI_smoluchowski_eqn}-\eqref{eq:SI_coordinate_conversion} is given by
\begin{equation}
	\frac{\mathrm{d} \bm{r}_{i} (t)}{\mathrm{d} t}
	=
	-\bm{u}(t)
	-
	\frac{1}{\gamma} \bm{\nabla}_{\bm{r}} V[\bm{r}_{i}(t)]
	+
	\sqrt{\frac{2kT}{\gamma}}\bm{B}_{i}(t)
	,
	\qquad
	i = 1, 2, \dots, N_{\text{p}}
	,
	\label{eq:SI_langevin_eqn}
\end{equation}
where $i$ is the particle index, $N_{\text{p}}$ is the total number of particles in the system, and $\bm{B}_{i} (t)$ is a white-noise source with statistics,
\begin{equation}
	\braket{\bm{B}_{i}(t)}
	=
	\bm{0},
	\qquad
	\braket{\bm{B}_{i}(t)\bm{B}_{i}(t')}
	=
	\delta(t - t')
	\tens{I}
	.
	\label{eq:SI_fluctuation_dissipation_theorem}
\end{equation}
[Note that the angle brackets $\langle\,\cdot\,\rangle$ appearing in Eq.~\eqref{eq:SI_fluctuation_dissipation_theorem} denote {\it ensemble} averages and are not to be confused with the {\it cell} average defined in the main text.]
The potential-energy field $V(\bm{r})$ and convective velocity $\bm{u}(t)$ appearing in Eq.~\eqref{eq:SI_langevin_eqn} are given by Eqs.~\eqref{eq:harmonic_potential} and \eqref{eq:trap_velocity}, respectively. Interactions between particles have been neglected, so the $N_{\text{p}}$ equations of motion are uncoupled.
For the purpose of numerically time-advancing Eq.~\eqref{eq:SI_langevin_eqn}, it is convenient to shift to the laboratory frame in which the position of each particle is measured as $\bar{\bm{r}}_{i}(t) = \bm{r}_{0}(t) + \bm{r}_{i}(t)$, where $\bm{r}_{0}(t) = \int_{0}^{t} \bm{u} (\tau) \, \mathrm{d} \tau = \hat{\bm{e}}_{x} A \sin{(\omega t)}$ denotes the time-dependent position of the moving traps. In this frame, Eq.~\eqref{eq:SI_langevin_eqn} becomes
\begin{equation}
	\frac{\mathrm{d} \bar{\bm{r}}_{i}(t)}{\mathrm{d} t}
	=
	-
	\frac{1}{\gamma} \bm{\nabla}_{\bar{\bm{r}}} V[\bar{\bm{r}}_{i}(t)-\bm{r}_{0}(t)]
	+
	\sqrt{\frac{2kT}{\gamma}}\bm{B}_{i}(t)
	,
	\qquad
	i = 1, 2, \dots, N_{\text{p}}
	.
	\label{eq:SI_langevin_eqn_relative}
\end{equation}
Here, the convective term has been eliminated and the potential-energy field oscillates in time.

\begin{figure}[!h]
	\begin{center}
\includegraphics[width=1\linewidth]{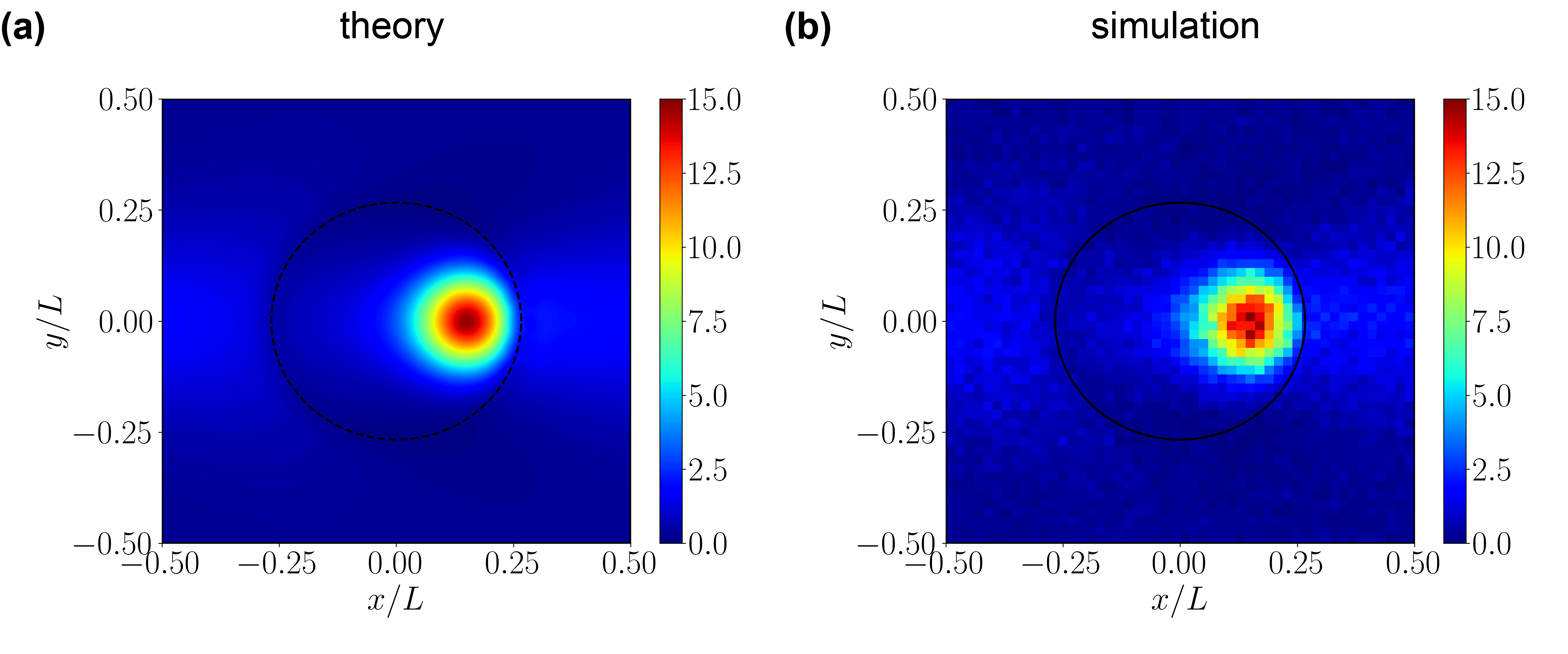}
		\caption{%
		Comparison of the convected probability density $g(x,y,t)$ for a stiff trap near the critical frequency ($\kappa=5$ $kT/\mu$m$^2$, $\omega/2\uppi=18.33$ mHz) from
		(a) deterministic solution of the Smoluchowski equation [Eq.~\eqref{eq:g_eqn}] and (b) stochastic simulation of the Langevin equation [Eq.~\eqref{eq:SI_langevin_eqn}].
		}
		\label{Fig:SI_Gfield_Theory_vs_Simulation}
	\end{center}
	\vspace{-18pt}
\end{figure}

In our Brownian dynamics simulations, we numerically advanced Eq.~\eqref{eq:SI_langevin_eqn_relative} using the GPU-enabled HOOMD-blue software package \cite{anderson2020hoomd}. A system of $N_{\text{p}} = 10,000$ particles was initialized at random positions within a periodically replicated $L\times L$ cell and advanced for $\tau = 10,000$ s (2.78 h) using a time step $\Delta t = 1$ ms. Fig.~\ref{Fig:SI_Gfield_Theory_vs_Simulation} shows that the simulated probability density shows excellent agreement with the deterministic solution of the corresponding Smoluchowski equation [Eq.~\eqref{eq:g_eqn}]. The MSD and effective diffusivity of the particles were then computed exactly as in the experiments using Eqs.~\eqref{eq:SI_msd_experiments}-\eqref{eq:SI_diffusivity_experiments}, wherein the time integral was discretized using the left Riemann sum.

\subsection{Derivation of Eq. (12): convective escape of a Brownian particle from a harmonic well}

We wish to estimate the critical oscillation frequency $\omega_{\text{max}}$ at which a Brownian particle rattling around the bottom of a potential-energy well is convected near the edge of the well with ample probability for escape. To make such an estimate, we start with the Langevin equation, Eq.~\eqref{eq:SI_langevin_eqn}, simplified for a single particle in a harmonic well $V(\bm{r}) = \tfrac{1}{2} \kappa r^{2}$:
\begin{equation}
	\frac{\mathrm{d} \bm{r} (t)}{\mathrm{d} t}
	=
	-
	\frac{\kappa}{\gamma} \bm{r}(t)
	-
	\bm{u}(t)
	+
	\sqrt{\frac{2kT}{\gamma}}\bm{B}(t)
	.
	\label{eq:SI_langevin_eqn_harmonic_well}
\end{equation}
Eq.~\eqref{eq:SI_langevin_eqn_harmonic_well} may be straightforwardly integrated with the initial condition $\bm{r}(0) = \bm{0}$ to give the fluctuating particle position,
\begin{flalign}
	\bm{r} (t)
	&=
	\mathrm{e}^{-\kappa t / \gamma}
	\int_{0}^{t}
	\mathrm{e}^{\kappa s / \gamma}
	\left(
		-\bm{u}(s)
		+
		\sqrt{\frac{2kT}{\gamma}} \bm{B}(s)
	\right)
	\mathrm{d} s
	.
	\label{eq:SI_langevin_eqn_harmonic_well_soln}
\end{flalign}
Substituting Eq.~\eqref{eq:trap_velocity} into \eqref{eq:SI_langevin_eqn_harmonic_well_soln} for the convective velocity then gives, upon integration,
\begin{flalign}
	\bm{r} (t)
	&=
	-\hat{\bm{e}}_{x} A 
	\left( \frac{\gamma \omega /\kappa}{1 + (\gamma \omega / \kappa)^{2}} \right)
	\left(
		\cos{(\omega t)}
		+
		\frac{\gamma \omega}{\kappa}
		\sin{(\omega t)}
		-
		\mathrm{e}^{-\kappa t / \gamma}
	\right)
	+
	\sqrt{\frac{2kT}{\gamma}} 
	\mathrm{e}^{-\kappa t / \gamma}
	\int_{0}^{t}
	\mathrm{e}^{\kappa s / \gamma}
	\bm{B}(s)
	\, \mathrm{d} s
	.
	\label{eq:SI_langevin_eqn_harmonic_well_soln_2}
\end{flalign}
The first term on the right-hand side of Eq.~\eqref{eq:SI_langevin_eqn_harmonic_well_soln_2} is the deterministic part of the fluctuating particle particle position, which is driven by oscillatory convection and attenuated by the trapping force. The second term is the stochastic part due to Brownian motion. The mean displacement and mean squared displacement of the particle respectively capture strength of these deterministic and stochastic elements:
\begin{flalign}
	\braket{\bm{r}(t)}
	=
	-\hat{\bm{e}}_{x} A 
	\left( \frac{\gamma \omega /\kappa}{1 + (\gamma \omega / \kappa)^{2}} \right)
	\left(
		\cos{(\omega t)}
		+
		\frac{\gamma \omega}{\kappa}
		\sin{(\omega t)}
		-
		\mathrm{e}^{-\kappa t / \gamma}
	\right)
	,
	\label{eq:SI_drift_in_harmonic_well}
\end{flalign}
\begin{flalign}
	\braket{(\bm{r}(t) - \braket{\bm{r}(t)})(\bm{r}(t) - \braket{\bm{r}(t)})}
	=
	\frac{kT}{\kappa} \left( 1 - \mathrm{e}^{-2\kappa t / \gamma} \right) \tens{I}
	,
	\label{eq:SI_variance_in_harmonic_well}
\end{flalign}
where we have applied the white-noise statistics, Eq.~\eqref{eq:SI_fluctuation_dissipation_theorem}, of the fluctuating $\bm{B}$-field.

After waiting a long enough time $t \gg \gamma /\kappa$, the exponential terms in Eqs.~\eqref{eq:SI_drift_in_harmonic_well}-\eqref{eq:SI_variance_in_harmonic_well} die off and we are left with an oscillating particle probability with variance $k T / \kappa$ given by Eq.~\eqref{eq:SI_variance_in_harmonic_well}. The amplitude of these oscillations are found from the extrema of the particle drift, Eq.~\eqref{eq:SI_drift_in_harmonic_well}:
\begin{equation}
	\sup_{t\ge 0}
	| \langle \bm{r}(t)\rangle |
	=
	\frac{\gamma \omega A/\kappa}{\sqrt{1+ (\gamma \omega / \kappa)^{2}}}
	\approx
	\frac{\gamma \omega A}{\kappa}
	\quad
	\text{for}
	\quad
	\frac{\gamma \omega}{\kappa} \ll 1
	.
\end{equation}
Thus, the basin of probability of size $\sim \sqrt{kT/\kappa}$ oscillates with amplitude $\sim \gamma \omega A / \kappa$ about the center of the potential-energy well. As the frequency $\omega$ is increased, the oscillations become more pronounced. The particle is expected to escape a well of finite width $W_{\text{trap}}$ when the spatial extent of the particle probability density crosses the edge of the well, at a critical frequency $\omega_{\text{max}}$:
\begin{equation}
	\tfrac{1}{2} W_{\text{trap}}
	\approx
	\frac{\gamma \omega_{\text{max}} A}{\kappa}
	+
	\sqrt{\frac{kT}{\kappa}}
	,
\end{equation}
or, solving for $\omega_{\text{max}}$,
\begin{equation}
	\omega_{\text{max}}
	\approx
	\frac{\kappa}{\gamma A}
	\left(
		\tfrac{1}{2} W_{\text{trap}}
		-
		\sqrt{\frac{kT}{\kappa}}
	\right)
	.
\end{equation}
The last expression is exactly Eq.~\eqref{eq:critical_frequency} from the main text.

\section{6. Additional Data}

\begin{figure}[!h]
	\begin{center}
		\includegraphics[width=0.5\linewidth]{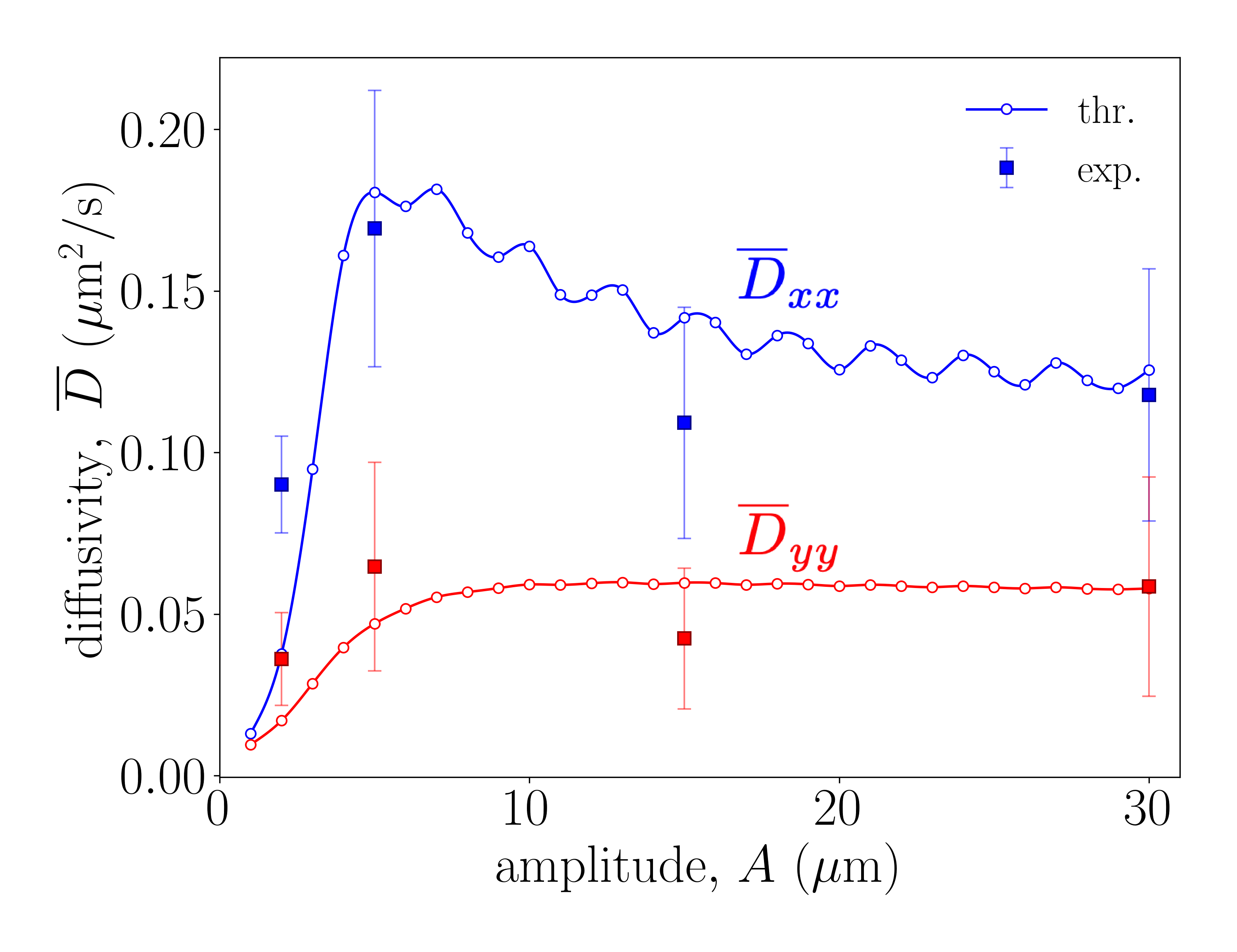}
		\caption{%
			Effective diffusivity as a function of oscillation amplitude at a fixed trap stiffness $\kappa = 5~kT / \mu\mathrm{m}^2$ and frequency $\omega/2\uppi = 18.3~\mathrm{mHz}$.
			Like Fig.~\ref{fig:Fig3} in the main text, $\overline{D}_{xx}$ is non-monotonic and reaches a maximum when the convection strength balances the harmonic trap strength.
			At very large amplitudes, the particle cannot quickly respond to the rapidly oscillating trap and explores regions outside of the harmonic well.
		}
		\label{Fig:SI4}
	\end{center}
	\vspace{-18pt}
\end{figure}

In addition to measuring the effective diffusivity $\overline{\tens{\bm{D}}}$ as a function of the oscillation frequency $\omega$, we also varied the amplitude $A$ while holding the frequency fixed.
The strength of the convective velocity $\bm{u}(t) = \hat{\bm{e}}_{x} \omega A \cos{(\omega t)}$ may be modified by varying either the amplitude $A$ or the frequency $\omega$. 
Fig.~\ref{Fig:SI4} plots $\overline{D}_{xx}$ and $\overline{D}_{yy}$ against $A$ for a fixed trap stiffness $\kappa = 5~kT / \mu\mathrm{m}^2$ and frequency $\omega/2\uppi = 18.3$ mHz.
This frequency corresponds to the critical frequency $\omega_{\text{max}}$ (for which $\overline{D}_{xx}$ is maximized) for $\kappa = 5~kT / \mu\mathrm{m}^2$ and $A = 5$ $\mu$m, as shown in the main text (see Fig.~\ref{fig:Fig3}).
We find that the $\overline{D}_{xx}$ is non-monotonic and achieves a maximum at $A = 5$ $\mu$m.
For amplitudes $A > 5$ $\mu$m, the convective motion is fast compared to the particle response time. Consequently, the particles sample regions outside of the harmonic well and their average diffusivity along the convection axis is reduced.

%
%

\pagebreak
\section{7. Supplemental Movies}
Below, we describe the Supplemental Movies associated with this manuscript. All time stamps corresponds to hours:minutes:seconds.

\begin{enumerate}
	\item[] \textbf{S1.} Experimental micrographs of silica particles with radius $a = 1.25~\mu$m diffusing through a stationary array of harmonic traps (6$\times$6 grid shown) with varying trap stiffness.
	
	\item[] \textbf{S2.} Microscopic Brownian dynamics simulations of a small sample of particles diffusing through a stationary array of harmonic traps (6$\times$6 grid shown) with varying trap stiffness (same parameters as in S1).

	\item[] \textbf{S3.} Experimental micrographs of silica particles with radius $a = 1.25~\mu$m diffusing through an oscillating array of stiff traps (6$\times$6 grid shown) with varying oscillation frequency and fixed trap stiffness $\kappa = 5~kT/\mu\mathrm{m}^2$. The second part of the movie shows the trajectories of several tagged particles.
	
	\item[] \textbf{S4.} Microscopic Brownian dynamics simulations of a small sample of particles diffusing through an oscillating array of stiff traps (6$\times$6 grid shown) with varying oscillation frequency and fixed trap stiffness $\kappa = 5~kT/\mu\mathrm{m}^2$ (same parameters as in S3).
	
	\item[] \textbf{S5.} Macroscopic Brownian dynamics simulations of 10,000 particles diffusing through a stationary array of harmonic traps (60$\times$60 grid shown) over long length and time scales, varying the trap stiffness.
	
	\item[] \textbf{S6.} Macroscopic Brownian dynamics simulations of 10,000 particles diffusing through an oscillating array of stiff traps (60$\times$60 grid shown) over long length and time scales, varying the oscillation frequency at a fixed trap stiffness $\kappa = 5~kT/\mu\mathrm{m}^2$.
	
	\item[] \textbf{S7.}  2D contour plots of the displacement field density $d_{x}(x,y,t)$ in an $L\times L$ periodic cell containing an oscillating harmonic trap, varying the oscillation frequency at a fixed trap stiffness $\kappa = 5~kT/\mu\mathrm{m}^2$ (same parameters as in S6). Bottom row plots the long-time average over one periodic cycle.
	
\end{enumerate}

\end{document}